\documentclass[twocolumn,superscriptaddress,floatfix,nofootinbib,prx]{revtex4}
\usepackage{graphicx,amsfonts,amssymb,amsmath,hyperref,hypcap,enumerate}
\usepackage{slashed}

\usepackage{amsmath,amssymb,bm,graphicx,dsfont}
\usepackage[latin9]{inputenc}
\usepackage{amsmath}
\usepackage{slashed}
\usepackage{dsfont}
\usepackage{amstext}
\usepackage{amssymb}
\usepackage{amsbsy}
\usepackage{amsthm}
\usepackage{epsfig}
\usepackage{graphicx}
\usepackage{bbm}
\usepackage{hyperref}
\usepackage{color}

\usepackage{slashed}

\newcommand{\bs}[1]{\boldsymbol{#1}}

\newcommand{\CT}{\mathcal{CT}}
\newcommand{\tCT}{\tilde{\mathcal{CT}}}

\newcommand{\be}{\begin{equation}}
\newcommand{\ee}{\end{equation}}
\newcommand{\bea}{\begin{eqnarray}}
\newcommand{\eea}{\end{eqnarray}}

\renewcommand{\vec}[1]{{\bf #1}}
\renewcommand{\hat}[1]{{\widehat #1}}

\begin{document}
\title{Composite fermion duality for  half-filled multicomponent Landau Levels}

\author{Inti Sodemann}
\affiliation{Department of Physics, Massachusetts Institute of Technology, Cambridge, Massachusetts 02139, USA}

\author{Itamar Kimchi}
\affiliation{Department of Physics, Massachusetts Institute of Technology, Cambridge, Massachusetts 02139, USA}

\author{Chong Wang}
\affiliation{Department of Physics, Harvard University, Cambridge, Massachusetts, 02138, USA}

\author{T. Senthil}
\affiliation{Department of Physics, Massachusetts Institute of Technology, Cambridge, Massachusetts 02139, USA}

\date{\today}
\begin{abstract}
We study the interplay of particle-hole symmetry and fermion-vortex duality in multicomponent half-filled Landau levels, such as quantum Hall gallium arsenide bilayers and graphene. For the $\nu{=}1/2{+}1/2$ bilayer, we show that particle-hole-symmetric interlayer Cooper pairing of composite fermions leads to precisely the same phase as the electron exciton condensate realized in experiments. This equivalence is easily understood by applying the recent Dirac fermion formulation of $\nu{=}1/2$ to two components. It can also be described by Halperin-Lee-Read composite fermions undergoing interlayer $p_x{+}ip_y$ pairing. An RG analysis showing strong instability to interlayer pairing at large separation $d\rightarrow \infty$ demonstrates that two initially-decoupled composite Fermi liquids can be smoothly tuned into the conventional bilayer exciton condensate without encountering a phase transition. We also discuss multicomponent systems relevant to graphene, derive related phases including a $Z_2$ gauge theory with spin-half visons, and argue for symmetry-enforced gaplessness under full SU$(N_f)$ flavor symmetry when the number of components $N_f$ is even.

\end{abstract}


\maketitle

\setlength{\pdfpagewidth}{8.5in}
\setlength{\pdfpageheight}{11in}

\section{Introduction}
In the last year, we have learnt of remarkable connections  between some seemingly distinct topics in quantum many body physics. Quantum Hall systems of two dimensional electrons in a half-filled Landau level have been related to correlated surface states of three dimensional topological insulators. The latter have in turn been related to three dimensional quantum spin liquid phases of insulating magnets. These connections have lead to a wealth of new insights and progress in all these research areas.  

On one end, it was conjectured by Son~\cite{Son15} that a simple way to reconcile the classic Halperin-Lee-Read theory~\cite{HLR} (HLR) of the compressible state that forms in a half-filled Landau level with particle-hole symmetry is by imagining that the composite fermion is a Dirac particle on which particle-hole acts effectively as time reversal.  On the other end, progress in understanding three dimensional time reversal symmetric quantum spin liquids  lead  to the discovery of a duality~\cite{Chong-dual,Chong-dual1,Max-dual} between the theory of a single Dirac cone (in $2+1$-dimensions) and a different Dirac theory where the Dirac fermions are coupled to a dynamical $U(1)$ gauge field.  Both theories arise at the surface of the same bulk three dimensional topological insulator (TI).  The duality interchanges the role of  time reversal symmetry ($U(1)\rtimes \mathcal{T}$) and  an anti-unitary charge conjugation ($U(1)\times \mathcal{C T}$). This duality is a generalization  of the particle-vortex duality, familiar in interacting bosonic $2+1$-d systems, to fermions.  This fermion-fermion duality clarifies a number of previously poorly understood issues  on the physics of symmetry enriched topological orders realized at the surface of fermionic topological insulators. Many aspects of the duality have since been further elaborated~\cite{CS16,wires,Max16,3dduality,karchtong,murugan,susydev,youxu}, and sharpened. 

In the quantum Hall context, the existence of such a dual description gives a theoretical basis to Son's proposed description of the half-filled Landau level. The Dirac composite fermions are simply understood as the dual fermions that arise in one side of the duality.  An intuitive physical picture of the Dirac composite fermion can also be developed as a a charge neutral composite of two $2\pi$ vortices bound to the electron carrying a finite dipole moment~\cite{CS16}.  The Dirac composite fermion theory finds further support in numerical calculations~\cite{DMRG}, and makes predictions for experiments~\cite{Son15,Tpower,CFsliquids} that might distinguish it from the HLR theory.  For further recent work on composite fermi liquids, see Refs.~\cite{Barkeshli,Kachru,MS16,Mulliganwires}.

Building on these developments, in this paper, we will revisit the physics of multicomponent  quantum Hall systems. These have been much studied over the years starting from work on bilayer quantum Hall systems and continuing to current work on graphene and related systems.  We will pay special attention to the role of particle-hole symmetry when it is present. Much of our focus will be on bilayer electronic quantum Hall systems at a total filling $\nu = 1/2 + 1/2$.  If the interlayer tunneling can be ignored, and for small interlayer separation, the system is in the celebrated exciton condensate phase~\cite{JimAllan,JimReview}. We will develop a new description of this state starting from  a `parent' compressible phase in which each layer has formed a composite fermi liquid. Along the way we will understand the action of particle-hole symmetry on the exciton condensate phase.  It has been known for a long time that the fundamental vortex defects around which the condensate order parameter winds by $2\pi$ carry fractional electric charge of $1/2$~\cite{Moon}. We will demonstrate that there exist vortex defects around which the order parameter winds by $4\pi$ which are charge neutral fermions, and moreover are Kramers doublets under the particle-hole symmetry. These neutral vortices are, as we show, the closest incarnation of the composite fermion itself in the exciton condensate.

Our treatment sheds new light on the old question of the fate of the quantum Hall bilayer at $\nu = 1/2 + 1/2$ as the interlayer separation $d$ is varied. What happens to the exciton condensate (known to be stable at small $d$) as $d$ is increased? At $d = \infty$ the two layers will be decoupled. Each layer is then expected to form a compressible composite Fermi liquid. As $d$ is decreased from $\infty$, it has long been recognized~\cite{Bonesteel}  that interlayer Coulomb interactions will lead to a pairing of the composite fermions. We will review this argument in a modern renormalization group framework (in Sec. \ref{pinst} below). The symmetry of the pairing channel is not determined by these calculations. We will obtain guidance from numerical work~\cite{Moller,Milo} that showed that interlayer composite fermion pairing in a $p_x + i p_y$ channel is energetically preferred. Remarkably, we find that  this $p_x + i p_y$ interlayer paired state is in the same phase as the exciton condensate that appears at small $d$.  Thus we are lead to a possible route for the evolution  from small to large $d$, which is simply that the exciton condensate is the ground state for all finite $d$. However we show that there will be some striking differences in some non-universal properties as $d$ is increased. We will see that at small $d$ the core energy for any vortex will be of order $e^2/l_B$  (where $l_B$ is the  magnetic length).  On the other hand, the pairing energy scale $\Delta$ for the composite fermions will go to zero as $d$ goes to $\infty$. In terms of the exciton condensate, we show that this implies that the core energy of the $4\pi$ vortex, which turns out to be controlled by $\sim \Delta$, is parametrically smaller than the core energy of the $2\pi$ vortex. This unusual phenomenon possibly can be detected in numerics/experiments in the future at moderately large-$d$.  We caution that the precise pairing symmetry of the composite fermions in the large-$d$ limit is hardly a settled issue. Indeed a very recent Eliashberg calculation~\cite{isobefu} found that a $p_x - ip_y$ channel is energetically favored in apparent disagreement with the numerical results in Refs.~\cite{Moller,Milo}. Additionally, other previous numerical studies have advocated for alternative phases to the exciton condensate beyond some critical $d$~\cite{Park,Yoshioka,PapicMilov}. We will not attempt to wade into this issue here. Though the pairing instability is itself a universal feature of the large-$d$ limit, it is likely that the pairing channel is sensitive to short distance physics. Our work is thus a demonstration that there need be no phase transition between the small and large-$d$ limits in some path in Hamiltonian space.

It is interesting to contemplate phases other than the exciton condensate that might be stabilized in this bilayer system. Indeed several such phases have already been proposed in the literature. As part of this paper we will address a specific related question. Is it possible to stabilize a gapped phase that preserves all the symmetries of the $\nu = 1/2 + 1/2$ bilayer in the lowest Landau level? Following discussions~\cite{Chong3D} (see also Ref.~\cite{Max3D})  of similar questions at the surface of the related $3d$ fermionic topological insulators with $U(1) \times \mathcal{CT}$, we will  construct a simple example of such a phase with a non-trivial topological order  described by a deconfined $Z_4$ gauge theory.  We conjecture that this is the simplest such symmetry preserving gapped state ({\em i.e} with the minimum number of topological quasiparticles).  The $Z_4$ topological order has a $16$-fold ground state degeneracy on a torus, and our conjecture implies that this is the minimum degeneracy of any symmetry preserving gapped state of the $1/2 + 1/2$ quantum Hall bilayer.

Additionally in this work we will also explore the cases of four- and eight-component half-filled Landau levels, exploiting their equivalence to the surface of chiral topological insulators (class AIII). In the presence of just the Coulomb interaction, the Hamiltonian of an $N$-component Landau level at half-filling will have $SU(N)$ symmetry in addition to charge-conservation and particle-hole symmetries. For $N$ even and a generic particle-hole symmetric Hamiltonian, we will provide a general argument for the impossibility of fully gapped topological order that preserves all symmetries. If some of the global $SU(N)$ symmetry is broken explicitly by the Hamiltonian, such a gapped symmetric topological order may be possible. We describe such topologically ordered states in some of these cases, obtaining them by quantum disordering broken symmetry states. This discussion essentially extends that of Refs.~\cite{Chong3D,Max3D}) by considering additional symmetries besides the microscopic particle-hole and electron number symmetries there described. One of our aims is to facilitate connections to realistic multicomponent systems, like graphene, where additional symmetries of the Hamiltonian might play an important role.

\section{Bilayer quantum Hall states at $\nu = \frac{1}{2} + \frac{1}{2}$}
\label{2lyrint}

Consider two quantum Hall layers each at filling $\nu = \frac{1}{2}$ with no interlayer tunneling.  This 
 physical situation is realized in spin-polarized gallium arsenide (GaAs) bilayers with negligible interlayer tunneling under a strong perpendicular magnetic field~\cite{JimReview}. This system can be described by a Hamiltonian projected to a single Landau level in which electrons $i$ and $j$ interact via two-body Coulomb potentials of the form:

\begin{equation}
\begin{split}
& V_{i j}=V_0 \left(r_i-r_j\right)+\tau _i^z \tau _j^z V_z \left(r_i-r_j\right),\\
& V_0 \left(r_i-r_j\right)+V_z \left(r_i-r_j\right)=\frac{e^2}{\epsilon \left|r_i-r_j\right|},\\
& V_0 \left(r_i-r_j\right)-V_z \left(r_i-r_j\right)=\frac{e^2}{\epsilon \left|d \hat{e}_z+r_i-r_j\right|},
\end{split}
\end{equation}

\noindent where $d$ is the distance separating the layers, and $\tau$ are Pauli matrices in the layer index space.  There are a number of global symmetries of this Hamiltonian that are important. First there are two $U(1)$ symmetries - which we denote $U_1(1)$ and $U_2(1)$ - associated with the conservation of the numbers $N_1$, $N_2$ of electrons in the top and bottom layers separately. In the limit when $d = 0$, the Hamiltonian is actually $SU(2)$ symmetric under rotations in layer space (known as `pseudospin').  This is broken to $U_1(1) \times U_2(1)$ at non-zero $d$. It will sometimes be convenient to consider the total charge $N_{+}  = N_1 + N_2$ and the ``pseudospin" $N_{-} = N_1 - N_2$.

Next, there is an antiunitary particle-hole symmetry - denoted\footnote{In the recent literature the same symmetry has also variously been denoted $C$ or $\mathcal{PH}$.} $\mathcal{CT}$ - which interchanges empty and full Landau levels of the bilayer system.  If we call the deviation from half-filling of the density of each layer as $\delta \rho_i \equiv \rho_i - \frac{B}{4\pi}$, then we have 
\begin{equation}
\mathcal{CT} \delta \rho_i (\mathcal{CT})^{-1} = - \delta \rho_i.
\end{equation}
Note that at $d = \infty$ the two layers are decoupled and we can do a particle-hole transformation separately for each layer. However at non-zero $d$ only the common $\mathcal{CT}$ operation is a symmetry. 

Finally there is an interlayer exchange symmetry $X$ which is unitary and simply exchanges the layer index.  If we call $\delta \rho_{\pm} = \delta \rho _1 \pm \delta \rho_2$, then 
\begin{eqnarray}
\label{rhopmCT}
\mathcal{CT} \delta \rho_{\pm} (\mathcal{CT})^{-1} & = &  - \delta \rho_{\pm}, \\
X\delta \rho_{\pm} X^{-1}  & = & \pm \delta \rho_{\pm}.
\label{rhopmX}
\end{eqnarray}

It is useful to consider the symmetries of the interlayer tunneling operator 
\begin{equation}
\label{Htunn}
H_{tunn} = - \sum_{a,b} c^\dagger_a (\bs{t}_\perp \cdot \bs{\tau})_{ab} c_b,
\end{equation}
where ${a,b}\in \{1,2\}$ are labels for the electron operators in either layer, and $\bs{t}_\perp = t_\perp (\cos \theta, \sin \theta, 0)$ is a vector in ``layer space" with components  only in the xy   ``pseudospin" plane. 
This is invariant under a diagonal subgroup of $U_1(1) \times U_2(1)$ (corresponding to conservation of the total charge $N_+$). Under $\CT$, $t_\perp \rightarrow -t_\perp$.   However $H_{tunn}$ is invariant under a modified anti-unitary particle-hole operation $\tilde{\mathcal{CT}} = \mathcal{CT} U_{1}(\frac{\pi}{2})U_2(- \frac{\pi}{2})$. One can choose the action of layer exchange, $X$, as $X c_a X^{-1}=(\bs{\hat{t}_\perp} \cdot \tau)_{ab} c_b$, so that the tunneling remains invariant: $\bs{t}_\perp \rightarrow \bs{t}_\perp$.

Bilayer quantum Hall systems of this sort have been studied intensely over the years. In the  $d \rightarrow 0$ limit with full $SU(2)$ symmetry the ground state is a quantum Hall pseudospin ferromagnet~\cite{Sondhi}. When $d \neq 0$ but is small, there is ``easy-plane" anisotropy, and the pseudospin points in the $xy$ plane: this corresponds to an exciton condensate with spontaneous interlayer coherence~\cite{Moon}.  This exciton condensate is a quantum Hall state and has $\sigma_{xy} = 1$ for the total charge current.  In the $d \rightarrow 0$ limit, there are skyrmion defects in the pseudospin ferromagnetic order which carry electrical charge $N_+ = 1$~\cite{Sondhi}. For $d \neq 0$ these split into two meron-vortices which cary charge $N_+ = \frac{1}{2}$~\cite{Moon}. As usual isolated vortices cost logarithmically large energy. 

In the limit $d \rightarrow \infty$, each layer will form a compressible composite fermi liquid state. How does the system evolve from this limit to the exciton condensate that is obtained in the opposite limit? One of our goals in this paper is to  address this question using the low energy effective field theory of the composite fermi liquid state. In Sec.~\ref{pinst} below we will review  and bolster - within a modern renormalization group framework - old arguments showing that in the large-$d$ limit the composite fermi liquids are unstable to interlayer pairing of the composite fermions. The fate of the system is determined by the specific pairing symmetry. We will be guided by previous numerical studies of this problem showing that the composite fermions of the two layers like to form a `pair' condensate in the $p_x+ ip_y$ channel as the separation $d$ is decreased~\cite{Moller,Milo}. Interestingly, we will show the resultant paired state is smoothly connected to the exciton condensate described above.  We will show this both within the framework of the Dirac composite fermion theory and the HLR theory.  

One outcome of our analysis through the Dirac composite fermions will be  to elucidate the role of particle-hole symmetry on the exciton condensate which does not seem to have been discussed in the literature. The exciton condensate order parameter may be taken to be precisely the $e^{i\theta}$ in the interlayer tunneling  operator of Eqn. \ref{Htunn}. As described above when it acquires an expectation value $\mathcal{CT}$ is broken but $\tilde{\mathcal{CT}}$ is preserved, and the question of how the latter symmetry acts on the excitations is meaningful. 

Composite fermion pairing channels other than the one supported by the exact diagonalization work of Refs. \onlinecite{Moller,Milo} are also in principle possible. These alternative pairing channels will not preserve particle-hole symmetry.  We will not study these other states. For some prior work on an example of such a state, see Ref. \onlinecite{Kim}.

\subsection{Equivalence between exciton condensate and interlayer composite fermion paired state}
We begin with two decoupled compressible composite Fermi liquid phases that are obtained in the limit $d \rightarrow \infty$. Each such composite fermi liquid is described by an effective theory of composite fermions forming a Fermi surface that are coupled to a fluctuating $U(1)$ gauge field. The precise description is however different in the Dirac and HLR theories, and so we will consider the two theories separately. We will analyse an interlayer paired state that emerges out of this parent compressible state.  Numerical work shows that such a paired state --- in the $p_x + i p_y$ {\em i.e angular momentum $l_z = 1$} channel --- is indeed energetically favored~\cite{Moller,Milo}. For the discussion below, it is important rightaway to note that the labelling of the  angular momentum pairing channel is different for the Dirac and HLR theories. The $\pi$ Berry phase at the Fermi surface in the Dirac theory implies that angular momentum  $j_z$ pairing of Dirac composite fermions is equivalent to angular momentum $l_z=j_z +1$ pairing of HLR composite fermions. Therefore when we analyse the paired state below, we will consider $j_z = 0$ pairing in the Dirac theory and, correspondingly, $l_z = 1$ pairing in the HLR theory. We will see explicitly that they lead to equivalent states. Of course the role of $\tilde{\mathcal{CT}}$ symmetry is only manifest in the Dirac theory.

\subsubsection{Two-component dual Dirac picture}\label{Dirac}

The decoupled Dirac composite fermi liquid is described by two copies of the action proposed by  Son~\cite{Son15}, and takes the form\footnote{As emphasized in Ref. \onlinecite{3dduality}, strictly speaking this theory should be refined to properly take into account global restrictions coming from quantization of coefficients of all Chern-Simons terms, including those involving the external background $U(1)$ gauge fields.  Accordingly we should regard the above action as a short hand for the more precise version described in Ref. \onlinecite{3dduality}.  For the purposes of the present paper, this subtlety does not play a crucial role and it is sufficient to work with the simpler action below.  Using the more precise version does not modify our conclusions.
} 
\begin{equation}\label{LDirac}
\begin{split}
&\mathcal{L}=\sum_{I =1}^2\bar{\psi}_I(i \slashed{\partial}+\slashed{a}_{I})\psi_I+\frac{1}{4 \pi }A_{I} d a_I+\frac{1}{8\pi} A_Id A_I+\cdots,\\
\end{split}
\end{equation}

\noindent

Here $\psi_I$, $a_{I}$, and $A_{I}$ are the composite fermion field, the internal $u(1)$ gauge field, and the external probe gauge field $A_I$ in layer $I$~\footnote{Our convention is as follows: a Chern-Simons term for gauge fields $\alpha,\beta$ read as: $\alpha d \beta\equiv \epsilon^{\mu\nu\sigma}\alpha_\mu\partial_\nu\beta_\sigma$, $x^\mu=(t,\vec{x})$, $A^\mu=(\phi,\vec{A})$, $j^\mu=-\delta \mathcal{L}/\delta A_\mu=(\rho,\vec{j})$, $\gamma^\mu=(\sigma_y,-i\sigma_z,i\sigma_x)$. The corresponding massless Dirac Hamiltonian only involves the real-symmetric pauli matrices: $H_0=\psi^\dagger (p_x\sigma_x+p_y\sigma_z)\psi$.}. The $\cdots$ contain non-universal terms including Maxwell terms for $a_I$ and long-range Coulomb interactions between the layers of the form $1/2\int dr dr' j_{I 0}(r) v_{I I'}(r-r') j_{I' 0}(r')$. The particle-hole conjugation acts on each of the Dirac composite fermions as a time reversal operation~\cite{Son15,Max-dual,CS16}: 

\begin{equation}
\begin{split}
&(\mathcal{C T}) \psi_I (\mathcal{C T})^{-1}=i \sigma _y \psi_I,\\
&(\mathcal{C T}) a^0_I (\mathcal{C T})^{-1}=a^0_I, \\
& (\mathcal{C T}) a^i_I (\mathcal{C T})^{-1}=- a^i_I.\\
\end{split}
\end{equation}

 The layer exchange symmetry, $X$, can be taken to act simply as $X \psi_I X^{-1}=\tau^x_{I J} \psi_J$, $X a^\mu_I X^{-1}=\tau^x_{I J} a^\mu_J$, where $\tau$ denotes Pauli matrices acting on the layer indices $I, J$ (summation implied).
  There are two separate internal $U(1)$ gauge symmetries  which we denote $u_{1}(1), u_{2}(1)$ (not to be confused with the global $U(1)$ symmetries associated with the physical charge conservation of each layer). It will sometimes be convenient to define the symmetric and antisymmetric gauge fields $a^{\pm}_\mu=(a_{1 \mu}\pm a_{2 \mu})/2$. Notice that the flux of the $a_{I \mu}$ gauge fields has the meaning of physical electron charge~\cite{Son15}, and hence, we will need to keep careful track of the correct quantization of fluxes of the $a^{\pm}$ gauge fields when we work with them. The corresponding gauge symmetries will be denoted $u_{\pm}(1)$, and the gauge charges $q_{\pm}$. Note that $q_\pm = q_1 \pm q_2$. 

\begin{figure}[t]
	\begin{center}
		\includegraphics[width=3.4in]{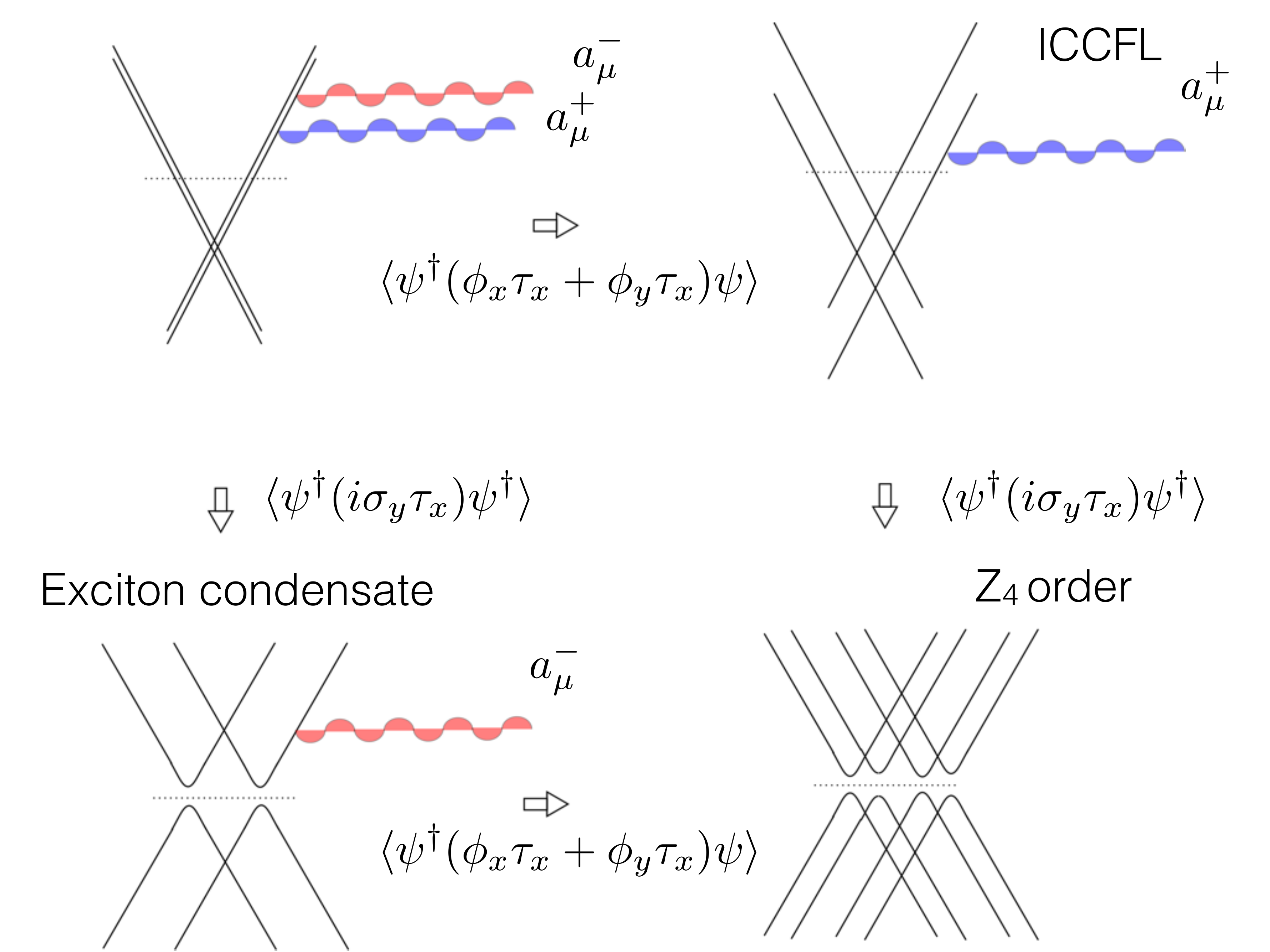}
	\end{center}
	\caption{(Color online) Schematic summary of the different two-component phases discussed in the main text. Starting from two composite fermion fermi seas coupled to two gauge fields (upper left) one obtains the conventional electron exciton condensate (lower left) via a particle-hole symmetric interlayer composite fermion pairing which higgses the layer symmetric $a^+_\mu$ gauge field, but leaves the layer asymmetric gauge field, $a^-_\mu$, gapless. Alternatively, one can induce an interlayer coherent composite fermion fermi liquid (ICCFL in upper right panel) via composite fermion exciton condensation, higgsing $a^-_\mu$ instead of $a^+_\mu$. If both condensations coexist one obtains a fully gapped particle-hole symmetric state with the topological order of $Z_4$ lattice gauge theory (bottom right).}
	\label{orders}
\end{figure}

We consider $j_z = 0$ pairing between the two species of composite fermions with the special property that  only the dual $u(1)_+$ is broken while the relative dual $u(1)_-$ is preserved. The specific form of inter-layer composite fermion pairing is:

\begin{multline}
\delta \mathcal{L}_\Delta= i g \Delta \psi \sigma _y \tau_x \psi -i g \Delta^* \psi^\dagger \sigma _y \tau_x \psi^\dagger\\
+|(i\partial_\mu+a_{1 \mu}+a_{2 \mu})\Delta|^2-u|\Delta|^2-\frac{v}{2}|\Delta|^4+\cdots
\end{multline}

\noindent Here $g$ controls the coupling of the fermions to the ``Cooper-pair"  field $\Delta$, and $u,v$ control the shape of the ``mexican-hat" potential dictating its condensation. Crucially, $\Delta$ has charge $q_+=-2$ under the symmetric gauge field $a^{+}_\mu=(a_{1 \mu}+a_{2 \mu})/2$   but it is neutral under the antisymmetric field $a^{-}_\mu=(a_{1 \mu}-a_{2 \mu})/2$.  Under $\mathcal{CT}$ and $X$, $\Delta$ transforms as
\be\label{CTXDelta}
(\mathcal{C T}) \Delta (\mathcal{C T})^{-1}=\Delta, \ X \Delta X^{-1}=\Delta.
\ee
\noindent Therefore, this dual ``superconductor" respects $\mathcal{CT}$, $X$, and the dual relative $u(1)_-$ gauge symmetry. Upon pairing, $\langle \Delta \rangle \neq0$, the Cooper-pair condensate will fully gap the gauge field $a^{+}_\mu$ via the Anderson-Higgs mechanism~\cite{Anderson}.  In terms of the physical electrons, this means that this phase is an electrical insulator. The flux of $a^+$ will be quantized to integer multiples  of $\pi$. These correspond physically to electrical charges $N_+$ that are quantized in units of $\frac{1}{2}$. 

However, the neutrality of $\Delta$ under $a^{-}$ implies that this gauge field is  not subject to an Anderson-Higgs mechanism. Additionally the $\mathcal{CT}$ invariance of the pairing guarantees that there will be no pairing-induced Chern-Simons term for $a^{-}_\mu$. As a consequence this gauge field is gapless and can be described at low energies by a pure Maxwell theory in two-dimensions.

A Maxwell theory for the $a^{-}_\mu$ gauge field coupled to the external probe gauge fields via a Chern-Simons term, as described in Eq.~\eqref{LDirac}, corresponds to the dual description~\cite{DH} of an electronic exciton condensate, as argued by Wen and Zee~\cite{WenZee}. Therefore our paired state of composite fermions is a condensate of interlayer excitons made out of the electrons, and has a spontaneously broken $U_-(1)$ symmetry, which is the  subgroup of $U_1(1) \times U_2(1)$  associated with the conservation of $N_{-} = N_1 - N_2$. It is \textit{a priori} conceivable, however, that this state does not describe the same   phase as the conventional exciton condensate described by a 111 Halperin-type wave-function, but could instead possesss distinct gapped quasiparticles. We will show, however, that the excitations of this paired state are in one-to-one correspondence with the topological defects and quasiparticles of the conventional exciton condensate~\cite{Moon}.

The gapped excitations of the paired state consist of Bogoliubov quasiparticles that descend from the composite fermions, and topological defects (vortices) of the pair condensate.  As is commonly done for superconductors~\cite{SF}, it is convenient to describe the Bogoliubov quasiparticles by stripping off their $a_+$ charge by writing $\psi_I = e^{i\frac{\phi_+}{2}} \epsilon_I$ (where the pair order parameter $\Delta \sim e^{i\phi_+}$). Though the fermions $\epsilon_I$ are neutral under $a_+$, they carry $a_-$ charges of $q_- = \pm 1$.  Further these symmetries allow $\epsilon_2$ to mix with $\epsilon_1^\dagger$ (as is explicitly seen by writing out the pairing term in terms of $\epsilon_I$).  Thus we will simply write these as $\epsilon_1 \sim \epsilon$, $\epsilon_2 \sim \epsilon^\dagger$. 

The topological defects (vortices)  of the superconducting paired condensate have winding of the phase of the pair field by $2 n\pi$, and associated quantized flux $n\pi$ of the internal gauge field $a_+$.  As mentioned above, these correspond physically to total electric charges $N_+ = \frac{n}{2}$.  To avoid confusion we emphasize that these are vortex defects of the pair field of the composite fermion, and not the vortices of the physical exciton condensate.  To distinguish these two we will label the former $n$-defects and use the term vortices exclusively for the latter. 

As we explicitly show in Appendix~\ref{dualBdG}, the BdG equations for these $n$-defects  are formally equivalent to those of the Fu-Kane superconductor~\cite{FuKane}, except that zero modes at odd-strength vortices correspond to full complex fermion zero modes (two Majorana modes). As a consequence $n$-defects with $n$ odd possess a zero complex fermion mode and for $n$ even, they do not. 

Consider first the $1$-defect. This has a single complex fermion zero mode, and consequently there are two such defects which we label $V_\pm$.  $V_+$ is obtained from $V_-$ by binding an $\epsilon$. 
This implies first that the $q_-$ charges of these two $1$-defects must differ by $1$.  Further under the layer exchange symmetry $X$,  $\epsilon \rightarrow \epsilon^\dagger$, as described in Appendix~\ref{dualBdG}. This in turn implies that $X$ interchanges $V_+$ with $V_-$.  Thus the consistent assignment of charge under $a^{-}_\mu$ for $\{V_+,V_- \}$ is $q_-=\{-1/2,1/2\}$ respectively.  
Second, from the definition of $\epsilon$, it is clear that when taken around these $1$-defects, there is a phase of $\pi$, {\em i.e}, they are mutual semions.  It follows that $V_+$ and $V_-$ are themselves also mutual semions. 

Next we turn to $2$-defects. These can be obtained as composites of the $1$-defects, {\em i.e} as $B_\pm = V_\pm^2, f ^\dagger= V_+ V_-$.  The $B_\pm$ have $q_- = \pm 1$ respectively while $f^\dagger$ has $q_- = 0$.  Note that $f^\dagger$ is a mutual semion with both $V_+$ and with $V_-$. In contrast $B_\pm$ are local around $V_\pm$. Since the layer exchange swaps $V_+$ and $V_-$, it follows that $f^\dagger$ maps onto itself under layer exchange. 
\footnote{
It is interesting to discuss the self-statistics of these defects. To do so we imagine temporarily ``turning off" the coupling to the fluctuating $a_-$ field.  Then $V_+, V_-$ can both be taken to be bosons.   The $B_\pm$ are bosons while $f^\dagger$ is a fermion. We can now formally introduce fields with these statistics, and couple them to $a_-$ according to their $q_-$ charges. }

$n$-defects with other values of $n$ may be discussed similarly. Let us now interpret these different excitations directly in terms of the electrons. We already pointed out that $n$-defects have electric charge $N_+ = n/2$.  Conversely a charge $q_{-}$ under the gauge field $a^{-}_\mu$ corresponds to vortices with $4 \pi q_{-}$ winding for the physical order parameter of the exciton condensate~\footnote{A simple way to elucidate this connection is to imagine gauging the external layer asymmetric probe gauge field $A^{-}_\mu=(A_{1 \mu}-A_{2 \mu})/2$ and noting that in this case the exciton condensate vortices in which the order parameter winds by $4 \pi q_{-}  $ would trap flux $2 \pi q_{-}$ of this gauge field.}.  Consequently, the $1$-defects $V_\pm$ carry physical charge $N_+=1/2$ and have vorticity $\pm 2\pi$ for the exciton order parameter. They thus correspond precisely to the vortex and anti-vortex meron defects of the exciton condensate with positive charge~\cite{Moon}.  The $2$-defects $B_\pm$ correspond to $4\pi$ vortices of the exciton order parameter with total charge $N_+ = 1$. 

More interesting are the two fermions $f^\dagger$ and $\epsilon$. As a $2$-defect $f^\dagger$ has $N_+ = 1$ but it has no vorticity.  Recall further that $f^\dagger$ is a mutual semion with the basic meron defects $V_\pm$.   These are exactly the same properties as the relic of the electron in the exciton condensate. Specifically we ``neutralize" the $A_-$ charge of the electrons in the top and bottom layers, $c^\dagger_1, c^\dagger_2$, by writing them as $c^\dagger_1 = e^{i\frac{\theta}{2}} f^\dagger_1, c^\dagger_2 = e^{-i\frac{\theta}{2}} f^\dagger_2$.  Now in the exciton condensate $f^\dagger_1$ and $f^\dagger_2$ can mix with each other and they count as a single common excitation $f^\dagger$ which has $N_+ = 1$, and which is a mutual semion around the basic $2\pi$ merons. 

The $\epsilon$ particle is a $0$-defect, and hence has $N_+ = 0$. However it carries $q_-= 1$, and hence is a $4\pi$ vortex of the exciton condensate. This electrically neutral $4\pi$ vortex can be obtained directly in the exciton condensate by binding a charge $1/2$ meron $V_+$ to a charge $- 1/2$ meron   denoted by $\bar{V}_-$. $\bar{V}_-$ is the antiparticle of $V_-$, carryinig $q_-=1/2$ and and hence $2\pi$ vorticity of the exciton condensate order parameter. $V_+$  can be obtained from $\bar{V}_-$ by binding with $f^\dagger$. As $f^\dagger$ is a mutual semion with both $V_+$, and $\bar{V}_-$, it follows that $V_+$, and $\bar{V}_-$ are themselves mutual semions. It is natural then that their bound state $\epsilon$ is a fermion.

This is exactly the same excitation structure as the usual exciton condensate. Thus, as promised, we learn that $j_z = 0$ interlayer pairing of Dirac composite fermions leads to a state that is smoothly connected to the usual exciton condensate. Further as we explain below we readily infer how the particle-hole symmetry acts on the exciton condensate which, to our knowledge, has not been discussed in the literature before.

We first recall that the relevant symmetry that is unbroken by the exciton condensate is the $\tilde{\mathcal{CT}}  = \mathcal{CT} U_{1}(\frac{\pi}{2})U_2(- \frac{\pi}{2})$ introduced above.  
The $f^\dagger$ particle is obtained from the electron by stripping off it's $U_-(1)$ charge. Thus it  transforms as: 
\be
\tCT f^\dagger {\tCT}^{-1} = f
\ee
We know that the phase $\theta$ of the exciton condensate is invariant under $\tilde{\mathcal{CT}}$, and hence the vorticity is left invariant by $\tilde{\mathcal{CT}}$ but their physical charge $N_+$, if any, will change sign. 
The merons $V_+$ and $\bar{V}_-$ are thus interchanged by $\tCT$.  Since they are mutual semions, their bound state - which is just $\epsilon$ - will be a Kramers doublet under $\tCT$. This also follows very directly from the Dirac composite fermion picture. As we have emphasized $\epsilon$ is simply the remnant of the composite fermion (which is a Kramers doublet under $\tCT$) in  the interlayer paired state. 

\begin{figure}[t]
	\begin{center}
		\includegraphics[width=3.2in]{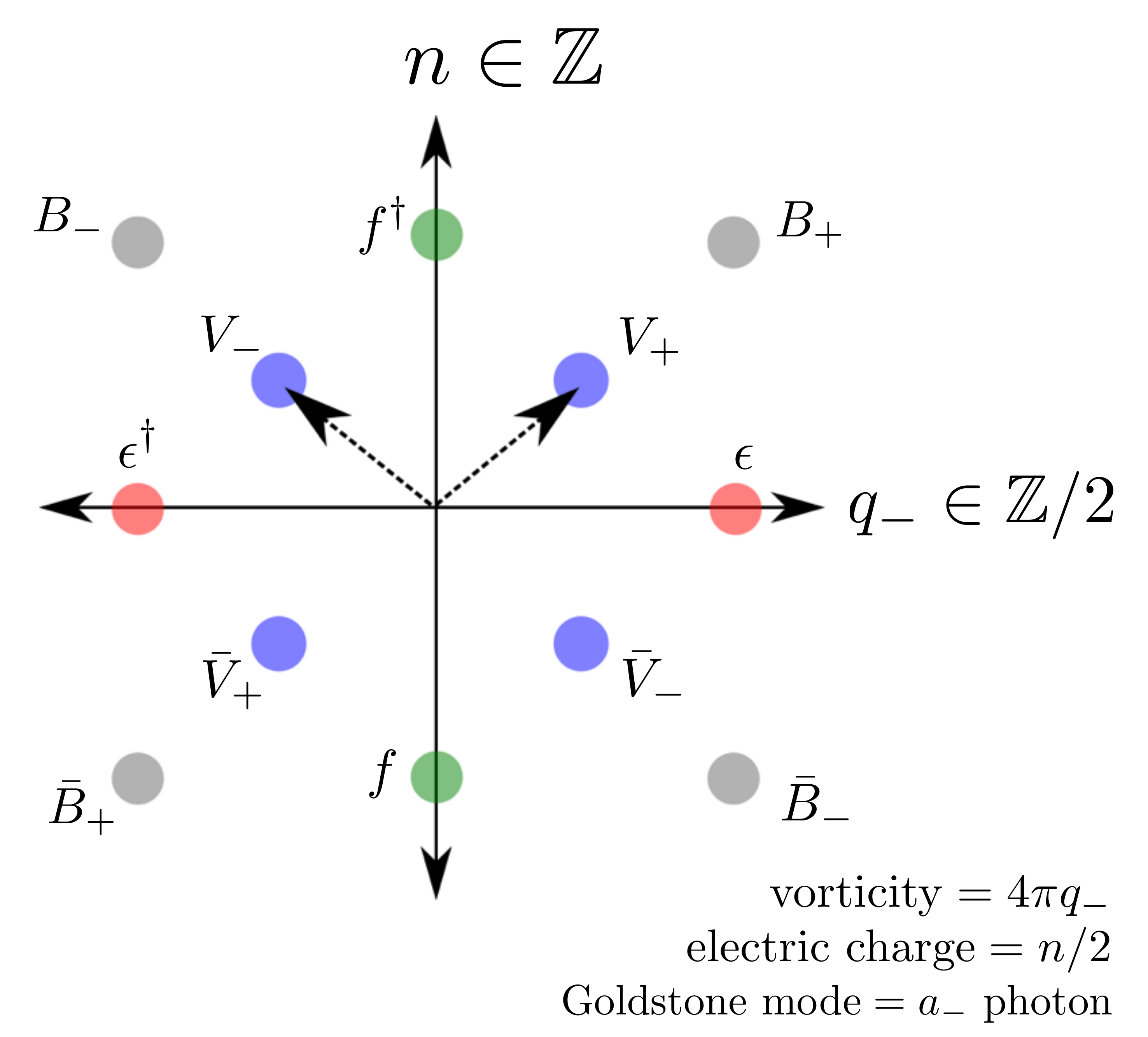}
	\end{center}
	\caption{(Color online) Quasiparticle lattice of the exciton condensate. The horizontal axis is the dual $q_-$ charge which is in one-to-one correspondence with the winding of the physical exciton order parameter, which is $4\pi q_-$. The blue dots designate the meron excitations. The vertical axis is the dual flux under $a_+$ in units of $\pi$ which is in one-to-one correspondence with the physical charge $N_+=n/2$. All quasiparticles can be constructed as bound states of elementary merons $V_+$ and $V_-$. Layer exchange symmetry $X$ acts as a mirror operation for the horizontal axis, $(q_-\rightarrow -q_-$, $n\rightarrow n)$, and the particle-hole symmetry $\tilde{\mathcal{CT}}$, acts as a mirror operation for the vertical axis , $(q_-\rightarrow q_-$, $n\rightarrow -n)$.}
	\label{excitonqps}
\end{figure}

The $\tCT$ transformation of other excitations can now readily be worked out. We have described $\tCT$ in the exciton condensate using its construction from the Dirac composite fermion theory. In Appendix~\ref{altCT} we give an alternate derivation of the $\tCT$ properties of this phase by constructing it directly in terms of electrons. The lattice of quasiparticles and the symmetry action is summarized in Fig.~\ref{excitonqps}. 

 We describe explicit wavefunctions for the  vortices in further detail in Appendix~\ref{Evortices}. It is interesting to note that configurations with only one vortex in the order parameter, of arbitrary vorticity, can be realized by having different magnetic field strengths for external magnetic fields acting on each of the two-components of interest, for example, in the sphere the exciton condensate ground state is realized at flux quanta $N^1_\phi=N^2_\phi=2 N_1-1=2 N_2-1$, with $N_{1,2}=N/2$, and a vortex of vorticity $v$ can be realized by setting instead $N^1_\phi=2 N_1-1+v$, $N^2_\phi=2 N_1-1$ with $N_{1,2}=N/2$, this can allow numerical studies of these vortices and verify their Kramers structure explicitly.

Let us comment on the issue of confinement. The vortices will, as usual, have logarithmic energy cost due to the phase winding of the order parameter. A weak  interlayer tunneling term, if present explicitly in the Hamiltonian, will pin the order parameter phase, and will lead to linear confinement of the vortices.

\subsubsection{Halperin-Lee-Read picture}\label{HLRpic}

In this section we consider the same paired state within the Halperin-Lee-Read (HLR) description. As already mentioned in HLR a pairing channel with orbital angular momentum channel $l_z$ corresponds to a spin-orbital coupled total angular momentum channel $j_z=l_z-1$ in Dirac picture~\cite{Son15}. Thus we study  $l_z=1$, or $p_x+ip_y$, in HLR picture~\footnote{  More precisely, the weak paring phase of $l_z=1$ channel in HLR picture corresponds to $j_z=0$ pairing in the Dirac picture.}. This is also a pseudospin triplet channel with $N_-=N_1-N_2=0$. Trial wavefunctions with this pairing symmetry have been shown to display large overlaps with ground states in exact diagonalization studies in the regime of intermediate layer separations $d\gtrsim l_B$~\cite{Moller,Milo}.

We consider a slightly modified version of the original Halperin-Lee-Read theory based on a parton construction instead of conventional flux binding (see, e.g.,  Ref.~\cite{Maissam}). Parton constructions have the formal advantage of making it easier to keep track of the normalization of the unit charges of the emergent gauge fields, simplifying the task of deriving a properly quantized $K$-matrix theory~\footnote{In this parton construction the physical electron operator in layer $I$ is the product of a boson and fermion: $c^\dagger_I=\psi^\dagger_I b^\dagger_I$. The fermion carries charge $-1$ under an emergent gauge field $a_I$, and the boson carries charge $+1$. Additionally the boson carries the full physical charge under the external probe field $A_I$ and forms a Laughlin state at $\nu_I=1/2$. The composite fermions $\psi_I$ do not experience a net magnetic field and hence they form a fermi sea type-state and are chosen as Galilean fermions in HLR. The gauge field $\alpha_I$ in Eq.~\eqref{LHLR} is the dual to the boson current of $b_I$.}. The Lagrangian describing this theory is:

\begin{equation}\label{LHLR}
\begin{split}
\mathcal{L}=&\sum_{I =1}^2 \psi_I^\dagger \left(i \partial_t+a_{I0}-\frac{(p+a_I)^2}{2m^*}\right) \psi_I\\
&-\frac{1}{2 \pi}\alpha_{I} d \alpha_I-\frac{1}{2\pi}(a+A)_I d\alpha_I+\cdots.\\
\end{split}
\end{equation}

\noindent where $\alpha_I$ is the field dual to the boson current, conventionally used in the Chern-Simons description of the Laughlin state. Upon pairing the composite fermions form a ``superconductor". The theory that ensues can be viewed as the gluing of a two-dimensional ``superconductor" to the Laughlin theory of bosons described by the Lagrangian in the second line of Eq.~\eqref{LHLR}. We willl proceed by first describing the ``superconductor", and, later on, we will glue it back together with the Laughlin bosons. The ``superconductor" is made from paired two-component non-relativistic fermions in a $p_x+ip_y$, $N_-=0$ layer pseudo-spin triplet channel, coupled to two internal gauge fields: $a_{1,2}$.  Such pairing can be formally induced by adding the following Lagrangian to Eq.~\eqref{LHLR}:

\begin{multline}
\delta \mathcal{L}_\Delta= g \Delta \psi \tau_x (p_x-i p_y) \psi + g \Delta^* \psi^\dagger \tau_x (p_x+i p_y)\psi^\dagger\\
+|(i\partial_\mu+a_{1 \mu}+a_{2 \mu})\Delta|^2-u|\Delta|^2-\frac{v}{2}|\Delta|^4+\cdots
\end{multline}

After the pairing we can describe the ``superconductor" as a product of a neutral sector and a charged sector under $a^{+}_\mu=(a_{1 \mu}+a_{2 \mu})/2$. The neutral sector corresponds to a familiar paired state  in two-dimensions: superfluid He-III in its A-phase with spin triplet $S_z=0$ pairing, where the spin of the Helium atoms plays the role of the layer pseudo-spin degree of freedom.  In modern language this corresponds to $\nu_{Kitaev}=2$ in the Kitaev classification~\cite{Kitaev16}. Such topological order can be described by a Chern-Simons theory with level $4$. Additionally, the quasiparticles in the neutral sector carry charges under the $a^{-}_\mu=(a_{1 \mu}-a_{2 \mu})/2$ gauge field.  In analogy with the Dirac case, the BdG equation describing the vortices of this $p_x+ip_y$ superfluid is formally equivalent to two copies of the spinless $p_x+ip_y$ superfluid studied by Read and Green~\cite{RG}, so that the zero modes correspond to complex fermion modes (two majorana modes). Therefore, we can write the corresponding Chern-Simons theory for the gauge fields describing the ``superconductor" as:

\begin{equation}\label{Lsc}
\begin{split}
&\mathcal{L}_{sc}=\frac{1}{\pi}\beta_-d\beta_- +\frac{1}{2\pi}\beta_- d(a_1-a_2)+\frac{1}{2\pi}\beta_+d(a_1+a_2).
\end{split}
\end{equation}

\noindent Let us describe the meaning of the different charges under the several gauge fields. A charge $l_-\in \mathbb{Z}$ under the gauge field $\beta_-$ labels the different quasiparticles of the neutral sector. We denote the  quasiparticles corresponding to the  the labels $\l_-=\{2,1,0,-1\}$ as $\{\mu,v,1,\bar{v}\}$ respectively. Label $\mu$ is a fermion, $\bar{v}$ is a vortex with its complex fermion zero mode empty, and $v$ is the corresponding vortex with its zero mode filled. Since the layer exchange symmetry, $X$, is also manifest in the HLR formulation, we can use the same argument employed in the Dirac case to infer the charge assignment for these vortices under $a^{-}_\mu$. Therefore $\l_-$ also determines the charge $q_-=-l_-/2$ under $a^{-}_\mu$. This is the physical origin of the mutual Chern-Simons term between $\beta_-$ and $a^-_{\mu}$ in Eq.~\eqref{Lsc}.

The charge $l_+\in \mathbb{Z}$ under $\beta_+$ labels the vortices of the charged sector, and the mutual Chern-Simons term between $\beta_+$ and $a^+_{\mu}$ encodes   the fact that Abrikosov vortices trap flux. $\beta_+$ is the dual field to the Cooper-pair current and the Chern-Simons term is familiar from the standard boson-vortex duality~\cite{DH}. We need to impose a further restriction on the allowed quasiparticles of this superconductor: the odd strength vortices in the charged sector have to appear in combination only with $\{v,\bar{v}\}$, and the even strength appear only with $\{1,\mu \}$, namely only quasiparticles satisfying $(l_-+l_+)/2\in \mathbb{Z}$ are allowed. This can be easily accomplished by a change of basis in the lattice of allowed charges of $\beta_{+/-}$, which can be implemented by redefining:

\begin{equation}\label{betas}
\begin{split}
&\beta_1=\beta_++\beta_-,\\
&\beta_2=\beta_+-\beta_-,
\end{split}
\end{equation}

\noindent and demanding that the corresponding charges $l_{1/2}=(l_+\pm l_-)/2$ be integers. We are now in a position to glue back this ``superconductor" to the Laughlin bosons appearing in Eq.~\eqref{LHLR}. This can be accomplished by noting that the $a_{1,2}$ fields appear linearly in the Chern-Simons action, so, one can integrate them out to obtain at low energies a constraint between the internal gauge fields of the ``superconductor" and the Laughlin bosonic fields. The constraint that follows is simply: $\beta_I=\alpha_I$, for $I=\{1,2\}$. Then the Chern-Simons part of the Lagrangian of our topological field theory can be written in the form of a K-matrix theory:

\begin{equation}\label{Kmat}
\begin{split}
&\mathcal{L}=\frac{1}{4\pi}\alpha^T K d\alpha-\frac{1}{2\pi}A^T  d\alpha+\cdots,\\
&K=-\begin{pmatrix}
1 & 1\\
1 & 1
\end{pmatrix}, \ 
\alpha=
\begin{pmatrix}
\alpha_1 \\
\alpha_2
\end{pmatrix}, \ 
A=
\begin{pmatrix}
A_1 \\
A_2
\end{pmatrix},
\end{split}
\end{equation}

\noindent which is the conventional Chern-Simons theory describing the 111 Halperin state~\cite{WenZee}. 

Let us close this section by contrasting the HLR and Dirac pictures of the exciton condensate as an interlayer paired state of composite fermions. In the HLR picture the existence of a gapless mode for the $a^-$ gauge field can be viewed as the result of a cancellation of two self Chern-Simons terms of apparent different origin: a ``background" Chern-Simons term arising from the flux binding and a Chern-Simons term induced by the specific $p_x+ip_y$ pairing channel under consideration. Notice that the cancelation occurs for this specific pairing channel, and would not occur if instead we had paired the composite fermions in the $p_x-ip_y$ channel, in which case a net self Chern-Simons term for $a^-$ would remain endowing the gauge field with a gap and hence not leading to a broken symmetry state in the physical electron degrees of freedom. Such $p_x-ip_y$ pairing had been considered in Ref~\cite{Kim} and was shown to lead to a $(3,3,-1)$ Halperin type state, which is clearly topologically distinct from the exciton condensate. 

The $p_x+ip_y$ channel in HLR corresponds in the Dirac picture to a pairing channel which manifestly respects the particle-hole symmetry of the bare electrons, because, such symmetry is implemented as a time reversal operation on the composite fermions. Given that the exciton condensate retains the $\tCT$ symmetry, the Dirac picture thus gives a simple route to reach it through composite fermion pairing in a way which manifestly preserves this symmetry.

\subsubsection{Equivalence from explicit wave-functions and connection to previous numerical studies}

Numerical studies have found that trial paired states of the type considered here have large overlaps with the exact ground state at intermediate interlayer distances $d\gtrsim l$~\cite{Moller,Milo}. In this subsection we show a way to re-write a trial paired wavefunction in a form that shows its exciton condensate correlations more explicitly. We begin by noting that in the symmetric gauge the canonical wave-function describing the exciton condensate can be written as a Halperin wave-function of the form:

\be
\Psi_{111}=\prod_{i<j}(z_i-z_j)\prod_{i<j}(w_i-w_j)\prod_{i,j}(z_i-w_j),
\ee

\noindent where we have written a wavefunction with a definite number of particles, $N/2$, in layer 1 (2) with coordinates $z_i$ ($w_i$) and we have omitted the ubiquitous exponential factors of the Lowest Landau level. On the other hand a trial wavefunction for the interlayer paired state can be motivated to be:
 
\begin{equation}\label{Psipair}
\begin{split}
&\Psi_{pair}=\frac{\prod_{i,j}|z_i-w_j|^m }{\prod_{i<j}|z_i-z_j|^n |w_i-w_j|^n} \times \\
&\det \left[\frac{1}{\bar{z}_i-\bar{w}_j}\right]  \prod_{i<j}(z_i-z_j)^2\prod_{i<j}(w_i-w_j)^2,
\end{split}
\end{equation} 
 
\noindent where a projection into the Lowest Landau level is implicit. The first product of factors that involves only the absolute values of interparticle distance is intended to be a variational factor that controls the probability amplitudes but not the phases of the wavefunction, we leave $n$ and $m$ as arbitrary parameters at this point. The prefactor $\det \left[\frac{1}{\bar{z}_i-\bar{w}_j}\right]$ describes a $p_x+ip_y$ BCS wavefunction for layer pseudo-spin triplet with $N_-=0$, and the Laughlin bosonic Jastrow factors describe the correlation-hole associated with intra-layer two-flux binding. The factor $\prod_{i,j}|z_i-w_j|^m$ also regularizes the probability of the wave-function as $w_j\rightarrow z_i$. Now, by making use of the Cauchy identity:

\be
\det \left[\frac{1}{\bar{z}_i-\bar{w}_j}\right]=\frac{1}{\prod_{i,j}(\bar{z}_i-\bar{w}_j)}\prod_{i<j}(\bar{z}_i-\bar{z}_j) \prod_{i<j}(\bar{w}_j-\bar{w}_i),
\ee

\noindent we can obtain the following expression for the paired wavefunction (up to an overall sign):

\be
\Psi_{pair}=\frac{\prod_{i<j}|z_i-z_j|^{2-n} |w_i-w_j|^{2-n}}{\prod_{i,j}|z_i-w_j|^{2-m}}\Psi_{111}.
\ee

\noindent The above relation shows that if we view a particle $z_1$ as an ``impurity" moving in a many-body sea of vortex-like objects described by $z_i$'s and $w_j$'s, it acquires the same phases in both wavefunctions when moving around those vortices, and the difference is only the probability amplitude with which it approaches the cores of those vortices, and also in the probability with which vortex cores approach each other. Such factors are non-universal and are dictated by the specific choice of wavefunction we made. It is non-trivial to elucidate the effect of the Lowest Landau Level projection on these wave-functions, but this rewriting is further strong evidence that the two wave-functions are specific realizations of the same underlying phase of matter.

Apart from the numerical studies described in Refs.~\cite{Moller,Milo}, that have motivated us to focus on the $p_x+ip_y$ interlayer channel, other numerical studies had previously encountered persistent aspects of exciton condensation physics to large interlayer distance. In particular, Ref.~\cite{MollerGoldstone} found numerical evidence of the persistence of the linearly dispersing Goldstone mode up to large interlayer distance. We also note that other numerical studies have advocated the possibility of the disappearance of the exciton condensate at interlayer distances beyond a critical value~\cite{Park,Yoshioka,PapicMilov}. We hope that our work motivates further numerical studies that attempt to settle the nature of the ground state at intermediate interlayer distances.

\subsection{Pairing instability}\label{pinst}

It is of course an energetic question whether any composite fermion pairing at all occurs  at large $d$, and if so in which channel. Here we address the following question. In the limit of very large-$d$, is the state with two decoupled composite fermi liquids stable to the weak interlayer Coulomb interaction? Previously this was addressed by Ref.~\cite{Bonesteel}. Here we revisit this issue within the framework of modern renormalization group treatments of such questions~\cite{MetlitskiSenthil2015}. We will see that arbitrary weak interlayer Coulomb interactions causes a pairing instability. This calculation however cannot determine the specific pairing channel.

 The composite Fermi liquid bilayer is unstable to interlayer pairing through a mechanism dependent on the asymmetric gauge field $a^-$. First, in contrast to  the long-ranged Coulomb kernel of  the total-charge gauge field $a^+$, the asymmetric $a^-$ gauge field couples to imbalances of the layer charges, and so its Lagrangian has  a shorter-ranged kernel, associated with the dipole coupling between regions of local charge imbalance. Therefore, $a^-$ loses the fluctuation-stabilizing effects of the long ranged Coulomb interaction, instead fluctuating quite freely and providing non-Fermi-liquid behavior. Second, composite fermions on the two layers couple to $a^-$ with opposite charges, resulting in Amperian attraction rather than repulsion, and hence a strongly enhanced pairing.

To set up the renormalization computation, let us study these two ingredients  in more detail. 
First we observe the propagator for each of the two different combinations of gauge fields, which is set by the microscopic interactions between electrons. 
Because of flux attachment, in either the HLR or Dirac pictures, the electron Coulomb interaction Eq. 1 can be rewritten as a long range interaction between fluxes of the gauge field. The $a^+$, $a^-$ gauge field self-interactions are 
\begin{align}
L_a & = \sum_{I,J} (\nabla{\times} a_I)[r_i]  \frac{e^2 / \epsilon}{ \sqrt{(1{-}\delta_{I J})d^2 + (r_i-r_j)^2}} (\nabla{\times} a_J)[r_j]
\nonumber\\
& = (\nabla{\times} a^+)[r_i]  \frac{4 e^2}{ \epsilon |r_i-r_j|} (\nabla{\times} a^+)[r_j]
\nonumber\\
& \quad + (\nabla{\times} a^-)[r_i]  2f_d[r_i-r_j]\frac{e^2 }{\epsilon} (\nabla{\times} a^-)[r_j]
\\
& \qquad f_d[r]  \equiv \frac{1}{|r|} - \frac{1}{\sqrt{d^2 + r^2}} \rightarrow \frac{d^2}{ 2}\frac{1}{ |r|^3}
\end{align}
The long range $1/r$ Coulomb interaction can stabilize the gauge field against strong fluctuations, allowing the composite fermions to be preserved as long lived quasiparticles with a sharp Fermi surface, albeit with  corrections expected for a marginal Fermi liquid. However here only $a^+$ enjoys a long ranged Coulomb interaction; in contrast, $a^-$ couples to electric dipoles rather than electric charges, and therefore exhibits a short ranged interaction, decaying as $1/r^3$ at large distances.

Second, we may observe the form of the coupling between the composite fermions of each layer and the two different combinations of gauge fields. 
The coupling of the Fermi surface to the gauge field can first be written as a sum over decoupled patch actions, each describing a pair of opposite (antipodal) patches of the Fermi surface. Let $x$ be the axis separating the patches, ie $\hat{x}$ is normal to the two patches. The lagrangian for each such pair of antipodal patches, denoted as $\eta=\pm$, contains the following coupling between fermions and gauge field,
\begin{align}
L_c &= \sum_{\eta=\pm} \sum_{I=1}^2  \eta v_F \psi^\dagger_{I \eta} (-i \partial_x + a_{I x}) \psi_{I \eta}
\\
&= \sum_{\eta=\pm} \eta v_F \Big[ \sum_{I=1}^2   \psi^\dagger_{I \eta} (-i \partial_x ) \psi_{I \eta}
\nonumber\\
&  \qquad \qquad \quad
+a_x^+ \psi^\dagger_{1 \eta} \psi_{1 \eta} 
+ a_x^+ \psi^\dagger_{2 \eta} \psi_{2 \eta}
\nonumber\\
&  \qquad \qquad \quad
+ a_x^- \psi^\dagger_{1 \eta} \psi_{1 \eta} 
- a_x^- \psi^\dagger_{2 \eta} \psi_{2 \eta}
 \Big]
\end{align}
Observe that the two layers couple with the same charge to the symmetric gauge field, but with opposite charges to the antisymmetric gauge field.

The RG procedure can then be performed following Ref.~\onlinecite{MetlitskiSenthil2015}. The full action is composed of the Lagrangians $L_a$ and $L_c$ plus the remaining fermion kinetic term $L_f = \sum_{\eta,I}  \psi^\dagger_{I \eta} ( \partial_\tau - v_F \partial_y^2 / 2K) \psi_{I \eta}$ with $K$ the Fermi surface curvature. 
$L_a$ may be written for the patch theory with renormalized couplings $\tilde{g}$ as $L_{a} \sim  |q| (a_x^+)^2 / \tilde{g}^+ +  q^2 (a_x^-)^2 / \tilde{g}^-$. Here $\tilde{g}^-$, associated with the bare $|q|^3$ interlayer interaction, captures the analytic $q^2$ term which is automatically generated by $|q|^3$ and is more relevant; its UV value is  $\tilde{g}^-  \approx l_B/d$. 
The resulting $a^+$ and $a^-$ static gauge field propagators have the form $D^+(0,q) \sim \tilde{g}^+/|q|$ and $D^-(0,q) \sim \tilde{g}^-/q^2$.  
[See Ref.~\onlinecite{MetlitskiSenthil2015} for intermediate-energy constants; $\tilde{\alpha}$ there is $\tilde{g}$ here.]
The effective couplings $\tilde{g}^\pm$ between the fermions and the $a^\pm$  gauge fields are found to obey the following RG flow equations, 
\begin{align}
\frac{d\tilde{g}^+}{d\ell} &= - (\tilde{g}^+)^2
\\
\frac{d\tilde{g}^-}{d\ell} &= \frac{1}{2}\tilde{g}^- - (\tilde{g}^-)^2
\end{align}
The coupling $\tilde{g}$ is proportional to the square of the gauge charge, so $\tilde{g^\pm}\geq 0$, and $\tilde{g}^+$ flows to zero logarithmically, controlled by the long range Coulomb interactions. In contrast, the coupling to the strongly-fluctuating mode $a^-$ flows to a finite value $\tilde{g}^-_* = 1/2$.

Now we may consider the flow of the four-fermion BCS scattering vertex $V$, which couples different patches of the fermi surfaces. 
The BCS four-fermion vertex $V$ can be decomposed into angular momentum pairing channels $V_m$; each is found to show the same independent flow. 
Denote the BCS pairing as $V^+$ for pairing between patches in the same layer, and $V^-$ for pairing between patches in opposite layers. The pairing RG flow equations are found to be
\begin{align}
\frac{dV^+}{d\ell} &=  - (V^+)^2 + \tilde{g}^+ +  \tilde{g}^-
\\
\frac{dV^-}{d\ell} &=  - (V^-)^2 + \tilde{g}^+ -  \tilde{g}^-
\end{align}
The first term in each line is the usual Fermi liquid result, while the $\tilde{g}$ terms arise from the Amperian repulsion/attraction set by the relative sign of the fermion charges under the gauge field. 
Since the gauge coupling fixed point is at $\tilde{g}^+_* = 0$ but $\tilde{g}^-_* = 1/2$, it is clear that the intra-layer pairing $V^+$ flows to repulsive interaction, while the inter-layer pairing $V^-$ flows to attractive interactions, enforcing an instability to an inter-layer paired  state at low temperature.

There are two things to note about this instability. First, the RG flow does not determine the pairing form of the instability, since all interlayer channels have the same diverging flow; rather, the pairing channel of the strongest instability is determined by short distance physics. 
Second, we note that the short ranged $1/r^3$ bare form of the $a^-$ interaction leads to uncontrolled non-Fermi-liquid physics. Indeed, the RG procedure can be controlled~\cite{MetlitskiSenthil2015} by a double expansion~\cite{mrossde} in the number of fermion species $N$ and in the range $\epsilon$ of the $a^-$ interaction written as $1/r^{1+\epsilon}$. The temperature scale at which the non-Fermi-liquid physics is expected to be seen is here nominally of the same scale as the pairing instability gap, even at large-$N$. However, the $a^-$  kernel can be modified by hand to $1/r^{1+\epsilon}$ form, furnishing a control parameter $\epsilon$; in the small-$\epsilon$ regime where non-Fermi-liquid physics is controlled, the pairing instability is unavoidable and preempts destruction of the Fermi surface. This qualitative behavior may then be expected to carry over to the present uncontrolled case, leading to a pairing instability at any $d$. We also observe that while the result for strong interlayer pairing instability is strictly robust only within the double expansion, the intuition associated with the two necessary ingredients (opposite $a^-$ charges and strong $a^-$ fluctuations) suggests that at $\epsilon\sim N \sim 1$ the interlayer pairing channel will still show a strong instability.

\subsection{Some physical consequences} 

Let us now consider some of the physical properties of the exciton condensate as $d$ is decreased from $\infty$. The existence of a weak coupling instability to pairing in the $d\rightarrow \infty$ limit discussed in the previous section, and the numerical evidence for a ground state with large overlaps with interlayer paired trial wave-functions of the $p_x+ip_y$ type~\cite{Moller,Milo}, suggests the conjecture that the ground state for a disorder-free quantum Hall bilayer does not undergo a quantum phase transition at any finite $d$. It is possible, therefore, that there is only a smooth crossover from large to small $d$, though the non-universal physical properties could change quite significantly. A close analogy exists with the celebrated BCS-BEC crossover of fermions with an attractive interaction. The large-$d$ limit corresponds, in this analogy, to a BCS-like state of the composite fermions, and the small-$d$ limit to the BEC state. There are however some striking differences in the interpretation of some of the phenomenology.

\begin{figure}[b]
\includegraphics[width=0.9\columnwidth]{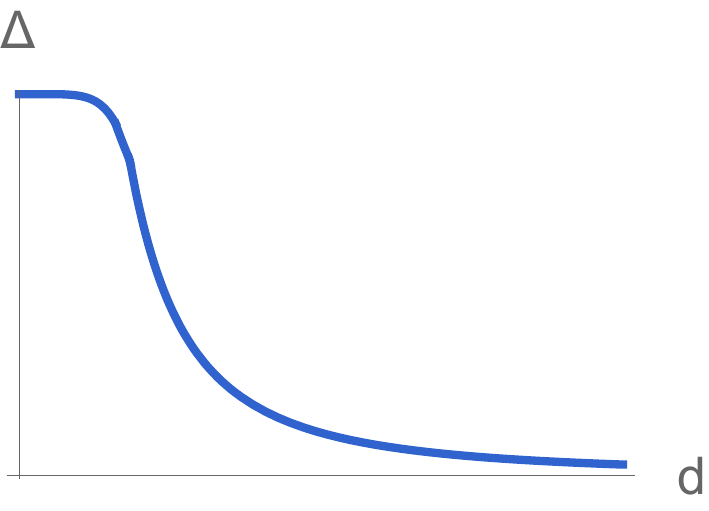}
\caption[]{(Color online) Schematic of the pairing gap $\Delta$ as a function of interlayer distance $d$. For large interlayer separations $d > l_B$, the pairing gap $\Delta$ (and the associated transition temperature) decrease rapidly as $\Delta \sim l_B^2/d^2$. For small $d$, the gap $\Delta$ approaches the Coulomb scale $e^2/l_B$. The gap $\Delta$ also sets the core energy of the $4\pi $ vortex; at large $d$, even fairly-separated pairs of $4\pi$ vortices will be cheaper than their $2\pi$ counterparts. } \label{fig:delta}
\end{figure}

In the large-$d$ limit, the pairing scale $\Delta$ (which also sets the energy scale for the  Berezinski-Kosterlitz-Thouless transition  out of the exciton condensate, $T_{BKT} \sim \Delta$ ) will be parametrically small in the ratio of the interlayer and intralayer Coulomb interactions, $d/l_B$ where $l_B$ is the magnetic length. Following the RG flow equations from the small UV value $\tilde{g}^-  \approx l_B/d$ at large $d$, one finds (in both the controlled small-$\epsilon$ regime as well as in the physical regime) that the resulting pairing gap $\Delta$ decreases with layer distance as $\Delta \sim (l_B^2/d^2) e^2/l_B$. In contrast, in the limit of very small $d$, the interlayer and intralayer Coulomb energy scales approach each other, and $\Delta \sim T_{BKT} \sim e^2/l_B$. This behavior of $\Delta$ is shown schematically in Figure~\ref{fig:delta}.

This has a very interesting consequence. Consider the large-$d$ limit. The composite fermion pairing scale $\Delta$ may be taken to be the energy gap of the $\epsilon$ particle within a mean-field description which ignores the coupling to $a^-$. But this particle has the physical interpretation of being a neutral $4\pi$ vortex in the exciton condensate.  The gap $\Delta$ is properly then viewed as the core energy of this vortex. The energy cost associated with the coupling to the $a^-$ is the usual logarithmic energy associated with the phase winding of the exciton condensate order parameter. The coefficient of this logarithm is proportional to the phase stiffness; since we are interested in its partially-renormalized value at the intermediate  scale $R$ associated with the vortex pair distance (see below), we may approximate it by its $T=0$ value even at temperatures nearly as high as the transition temperature. 
The  $T=0$ stiffness can be estimated in mean-field~\cite{Moon} to be $e^2/16\sqrt{2\pi} l_B$ at $d=0$, decreasing at larger $d$ as $e^2 l_B^2 /8 \pi d^3$. 
What about the $2\pi$ vortices? In the large-$d$ limit these are obtained as vortex defects (the $1$-defects) of the composite fermion pair order parameter. A standard argument shows that they will have core energy set by the composite fermion Fermi energy $\sim e^2/l_B$.  Thus the core energy of the neutral $4\pi$ vortices is much smaller than that of the $2\pi$ vortices in the large-$d$ limit. Of course as usual the phase winding energy is smaller, by a factor of 4, for the $2\pi$ vortices.

As $d$ decreases the core energy of the $4\pi$ vortex will increase and at some point become comparable to that of the $2\pi$ vortex ($C e^2/l_B$ with some numerical coefficient $C$). If we create vortex-antivortex pairs separated by a distance $R$, the total energy is the sum of core and phase winding contributions.  For sufficiently large $R$ at any $d$, $2\pi$ vortex-antivortex pairs will be cheaper than their $4\pi$ counterparts. However at large-$d$, there will be a range of $R \gg l_B$ where it will be cheaper to create $4\pi$ vortex-antivortex pairs than $2\pi$ ones. In the mean-field estimate this scale $R$ is exponentially large in $d^3$,  
\begin{align}
R &\approx l_B \exp\left[ \frac{C e^2/l_B - \Delta}{(2^2 - 1^2) \pi \rho_s}  \right] 
\approx l_B \exp\left[ \tilde{C}\frac{d^3}{l_B^3}
\right]
\end{align}
and can become effectively thermodynamically infinite at large $d$. 
This may reveal itself near the finite temperature phase transition, where the stiffness may be unusually renormalized as a function of temperature, when $d$ becomes sufficiently large such that $4\pi$ vortex pairs become cheaper than the fundamental $2\pi$ vortices. 

In particular, observe that at large $d$, there is a separation of energy scales $\pi \rho_s \ll \Delta \ll C e^2/l_B$ between the stiffness, the $4 \pi$ vortex core energy and the $2 \pi$ vortex core energy, respectively. Thus the finite temperature transition into the exciton condensate is expected to remain a continuous Kosterlitz-Thouless transition. The transition will still be associated with a discontinuity in the stiffness $\rho_s$ at $T_c$, jumping from $\rho_s=0$ up to the diagonal line $\rho_s = (2/\pi) T$. Here, however, as $\rho_s$ continues to rise with decreasing temperature, there is also expected to be a rounded singularity reaching up to the diagonal line $\rho_s = (8/\pi) T$, associated with the pairs of $4\pi$ vortices. This additional sharp rise in $\rho_s$ should occur at $T$ quite close to $T_c$. If the length scale $R$ above is increased (say by increasing $d$) up to the finite size $L$ of the mesoscopic system, the $8/\pi$ jump would become fully singular, and would occur precisely at $T_c$ through a double-vortex continuous Kosterlitz-Thouless transition.

These observations do not rely on the presence of particle-hole symmetry. If this symmetry is present,  as we argued, the neutral $4\pi$ vortex will be a Kramers doublet. It will be interesting to find a way to create and probe the associated two-fold degeneracy.  Our discussion ignores the impact of disorder in this phase. Numerical studies suggest that the exciton condensate is stable against weak quenched disorder and indicate the possibility of an intervening glassy phase at finite disorder strength in which the exciton superfluidity disappears but the system remains an insulator in the bulk with a quantized Hall conductivity before the system turns into a gapless state at stronger disorder~\cite{Sheng}.

\section{Symmetry respecting fully gapped $Z_4$ state at $\nu = \frac{1}{2} + \frac{1}{2}$}\label{Z4}

Though the exciton condensate is potentially a stable phase at any $d$, it is interesting to ask about other possible phases that might also be stable at intermediate or large $d$. Of particular interest to us here is the possibility of a gapped phase which preserves all the symmetries of the Hamiltonian. We will construct what we conjecture is the simplest example of such a phase and show that it has topological order described by a deconfined $Z_4$ gauge theory. This state has a $16$-fold topological ground state degeneracy on a torus, and our conjecture implies that this is the minimum possible degeneracy if all the symmetries are preserved.

This state has, to our knowledge, not been previously described within the quantum Hall literature but it is closely connected to the gapless exciton condensate state we just described and also to the Interlayer Coherent Composite Fermi Liquid (ICCFL) state proposed in Ref.~\cite{Alicea09}. We are interested in obtaining a fully gapped topologically ordered phase, therefore, we wish to gap out the $a^{-}_\mu$ photon, which is the linearly dispersing Goldstone mode viewed from the electron exciton condensate perspective. To achieve this while keeping the microscopic symmetries of the bare fermions we proceed by condensing a field that carries charge under $a^{-}_\mu$~\footnote{In the present work we describe the low energy theory of the resulting phase without focusing on the specific microscopics that might make it energetically favorable.}. In other words, in addition to the composite fermion cooper pair, we condense a dual exciton bosonic field, $\phi=\phi_x+i \phi_y$, made from a composite fermion particle-hole pair, which can be done by adding the following Lagrangian to Eq.~\eqref{LDirac} or Eq.~\eqref{LHLR}:
\begin{multline}\label{Lphi}
\delta L_\phi=g_\phi \bar{\psi} \gamma_0(\tau_x \phi_x+\tau_y \phi_y)  \psi\\
+|(i\partial_\mu+a_{1 \mu}-a_{2 \mu})\phi|^2-s|\phi|^2-\frac{r}{2}|\phi|^4+\cdots
\end{multline}
\noindent where $g_\phi,s,r$ are parameters controlling the coupling of $\phi$ to the fermions and its condensation. Under the particle-hole ($\mathcal{CT}$) and layer-exhange ($X$) symmetries, $\phi$ transforms as:
\be\label{CTXphi}
(\mathcal{C T}) \phi (\mathcal{C T})^{-1}=\phi, \ X \phi X^{-1}=\phi^*.
\ee
\noindent Therefore the condensation respects $\mathcal{CT}$. Since $\phi$ carries charge $2$ under  $a^{-}_\mu$ its condensation gaps the fluctuations of $a^{-}_\mu$ via the Anderson-Higgs mechanism~\cite{Anderson}. The interaction between $a^{-}_\mu$ charges becomes short-ranged and they become fully deconfined. The charge under $a^{-}_\mu$ in the exciton condensate picture implies that $\phi$ is an $8\pi$ vortex of the order parameter. Therefore this transition can also be viewed as a form of vortex condensation in a superfluid analogous to that occurring in the superconductor at the surface of a two-component AIII topological insulator described in Ref.~\cite{Chong3D}. In the resulting phase the fundamental vortex of the condensate of $\phi$ that traps $\pi$ flux of the $a^{-}_\mu$ gauge field at its core also becomes deconfined. This quasiparticle is charge neutral $N_+=0$ under the external symmetric probe gauge field $A^{+}_\mu$, and has fractional layer-charge-imbalance charge $N_-=1/2$ under $A^{-}_\mu=(A_{1 \mu}-A_{2 \mu})/2$. In the reminder of this section we will describe the properties of this phase both within the Dirac and HLR pictures.

\subsection{$Z_4$ ordered state from Dirac picture}\label{Z4Dirac}

We start from the dual description of the exciton condensate in the Dirac picture described in Section~\ref{Dirac}. We induce a phase transition starting from a parent exciton condensate by condensing an $8\pi$ vortex of the exciton order parameter which can be viewed as a pair of composite fermions $\phi \sim \epsilon^2$. This object carries charge $q_-=2$ under the $a^{-}_\mu$ field, and has no physical charge ({\it i.e.} it carries no flux under the $a^{+}_\mu$ field). This dual  exciton condensate field, $\phi$, must not be confused with the physical exciton order parameter which has disappeared in the present phase since the long-range order is destroyed by vortex condensation. The condensation allows for vortex-like topological defects of the dual exciton condensate, $\phi$, to become stable quasiparticles. These new defects can be labeled by an integer $m$. Around an $m$-defect the pair field $\phi$ has a phase winding of $2 m\pi$, and an associated quantized flux $m\pi$ of the internal gauge field $a^-$. As mentioned before, this flux corresponds physically to a total layer charge imbalance of $N_-= m/2$. Let us denote the elementary $m=\pm1$ defects $\beta$, and $\bar{\beta}$ respectively. $\bar{\beta}$ is the antiparticle of $\beta$ and both objects have bosonic self and mutual statistics. The $\beta$ vortex does not carry non-trivial zero modes, and hence its transformation properties under layer exchange, $X$, and particle-hole, $\mathcal{CT}$, symmetries follows simply from the fact that the $a^-$ flux is odd under either of these transformations and that the vorticity of  $\phi$ is also odd under these transformations (which follows from Eq.~\eqref{CTXphi}):

\be\label{CTXbeta}
\mathcal{C T} \beta \ \mathcal{C T}^{-1}\rightarrow \bar{\beta}, \ X \beta X^{-1}\rightarrow \bar{\beta}.
\ee

In this phase we have relics of the meron quasiparticles which we will denote by the same labels as in the exciton condensate $\{V_+,V_-,\bar{V}_+,\bar{V}_-\}$. Importantly, in the present phase the relics of the merons become fully deconfined finite energy excitations because their charge under the $a_-$ field is screened by the $\phi$ condensate. However, in spite of their $a_-$ charge being screened, they still experience a long-range statistical Aharonov-Bohm-type interaction with the vortices carrying $a_-$ flux, namely with the $\beta$ defects described in the previous paragraph, much like in the case of quasiparticles in superconductors. Consider for example $V_+$. Since this excitation carried $q_-=1/2$ under $a^{-}$ in the exciton condensate, this implies there is a phase of $\pi/2$ when it completes a full braid around $\beta$. Both of these quasiparticles have bosonic self-statistics and therefore they behave like the anyons of $Z_4$ gauge theory. As we will see, any other excitation can be expressed as a bound state of these two quasiparticles modulo local excitations, therefore the present state has indeed the topological order of $Z_4$ gauge theory.

Let us denote the physical electron quasiparticles in the top and bottom layers by $\{c_1^\dagger,c_2^\dagger\}$ respectively. $c^\dagger_a$ carries physical charges $N_+=1$ and $N_-=(-1)^{a+1}$. The quasiparticles $\beta^4$ and $V_+^4$ can be viewed as local bosons, indeed, $\beta^4\sim c^\dagger_1c_2$, and $V_+^4\sim c^\dagger_1c^\dagger_2$. Therefore, for purposes of describing the quasiparticles it will suffice to keep track of the electron operator only in one layer since $c_2^\dagger\sim \beta^4 c^\dagger_1$. The electron is local with respect to all excitations, and since $\bar{\beta}$ has bosonic self-statistics, it follows that $\bar{\beta}^2c^\dagger_1$ is a fermion. This fermion has a global physical charge $N_+=1$, but has no layer charge imbalance, $N_-=0$. Additionally, $\bar{\beta}^2c^\dagger_1$ acquires a phase of $-1$ when it completes a full braid around the $V_+$ meron, meaning that they have mutual semionic statistics. Therefore, this quasiparticle is the relic of the bogoliubov-like fermion of the exciton condensate, $f^\dagger$, described in Section~\ref{Dirac}. We will keep the same label for this quasiparticle in the $Z_4$ state, namely we call $f^\dagger= \bar{\beta}^2c^\dagger_1$. 

The identification $f^\dagger= \bar{\beta}^2c^\dagger_1$ allows to identify all the relics of the exciton condensate quasiparticles in the present order because they can be constructed as bound states of $\{V_+, \bar{V}_+,f^\dagger,f\}$. The relic of the meron with the same physical charge but with opposite vorticity of $V_+$, labeled $V_-$, can be obtained as $V_- = f^\dagger \bar{V}_+= \bar{\beta}^2  \bar{V}_+ c^\dagger_1$. The relic of the composite fermion particle $\epsilon$ can be obtained as $\epsilon = V_+ \bar{V}_-=\beta^2  V_+^2 c_1$, which can be seen to be a fermion ($\beta^2  V_+^2$ has bosonic self-statistics) and carries no physical charges $N_+=N_-=0$. From this identification we can specify how $V_+$ transforms under $\mathcal{CT}$ and $X$:

\begin{equation}\label{CTXV}
\begin{split}
&\mathcal{C T} \ V_+ \mathcal{C T}^{-1}\rightarrow \bar{V}_-= V_+ \beta^2 c_1,\\
&X \ V_+ X^{-1}\rightarrow V_- =\bar{V}_+ \bar{\beta}^2   c^\dagger_1.
\end{split}
\end{equation}

Notice that the since $V_+$ and $\bar{V}_-$ are swapped under $\mathcal{C T}$ and they are mutual semions, it follows that $\epsilon= V_+ \bar{V}_-$ is a Kramers fermion. This is consistent with the assigment in the exciton condensate $\tilde{\mathcal{CT}}^2=-1$, since $\tilde{\mathcal{CT}}^2=\mathcal{CT}^2$ for any quasiparticle that has no layer charge imbalance $N_-=0$.

\begin{table}
\caption{Topological spins of representative quasiparticles of the four distinct semion superselection sectors of $Z_4$ order in quantum Hall bilayers. Transformation rules under layer exchange $X$ and particle-hole symmetry $\mathcal{CT}$, and their total charge $N_+=N_1+N_2$ and layer charge difference $N_-=N_1-N_2$ are also listed. The electron operators in the two layers are related by $c^\dagger_2 \sim \beta^4 c^\dagger_1$.} 
\centering 
\begin{tabular}{c  c  c  c  c} 
\hline\hline 
 & $V_+ \beta$ & $V_+ \bar{\beta}$ & $\bar{V}_+ \bar{\beta}$ & $\bar{V}_+ \beta$ \\ [0.5ex] 
\hline 
$\theta$ & \ $-i$ \ & \ $i$ \ & $-i$ \ & $i$ \\ 
$X$ & \ $\bar{V}_+ \beta c_2^\dagger$ \ & \ $\bar{V}_+ \bar{\beta} c_1^\dagger$ \ & $V_+ \bar{\beta} c_2$ \ & $V_+ \beta c_1$ \\
$\mathcal{CT}$ & \ $V_+ \beta c_1$ \ & \ $V_+ \bar{\beta} c_2$ \ &  $\bar{V}_+ \bar{\beta} c_1^\dagger$ \ &  $\bar{V}_+ \beta c_2^\dagger$ \\
$N_+$ & \ $1/2$ \ & \ $1/2$ \ &  $-1/2$ \ &  $-1/2$ \\
$N_-$ & \ $1/2$ \ & \ $-1/2$ \ &  $-1/2$ \ &  $1/2$ \\ [1ex] 
\hline 
\end{tabular}
\label{tabZ4} 
\end{table}

A consequence of the anti-unitary nature of $\mathcal{C T}$ symmetry is that the super-selection sectors related by this symmetry must have topological spins which are complex conjugates of each other. $Z_4$ topological order contains only fermions, bosons, semions and anti-semions. Then, this rule implies that fermions map into fermions, bosons into bosons, but the semions must be mapped into anti-semions. Table~\ref{tabZ4} lists the topological spins and the transformation rules of representative semions belonging to the four distinct semion sectors of $Z_4$ order. It is noteworthy that $X$ exchanges topological sectors of the semions but leaves the topological spin invariant by binding a physical electron. On the other hand $\mathcal{CT}$ leaves the topological sectors invariant, but binds a physical electron to the semions changing them into anti-semions and viceversa. From this one can infer that the action of $\mathcal{CT}$ is to fill a single physical fermion zero mode, and, therefore it follows that $\mathcal{CT}^2$ is well defined on these semions and can be taken to be $\mathcal{CT}^2=1$.

\subsection{$Z_4$ ordered state from HLR picture}\label{Z4HLR}

A $K$-matrix theory for this phase can be obtained following a similar reasoning as in Section~\ref{HLRpic}. In this case we begin by striping-off the neutral sector entirely from its $a^{-}_\mu$ charges, by introducing a field, $\beta_-$, which is dual to the $\phi$ boson current and enforcing the corresponding Meissner effect for the vortices of such condensate, in exactly the same fashion as we did for the Cooper pair field in the charged sector of the superconductor. Therefore, instead of Eq.~\eqref{Lsc}, in the present case we write~\footnote{Another way to arrive at this Lagrangian is by writing the HLR composite fermion as the product of a fully neutral fermion, $\mu$, and bosons, $d_I$, which carry the $a_I$ charge: $\psi_I^\dagger=\mu \ d_I^\dagger$. $\mu$ and $d_I$ carry unit charges under an internal $Z_2$ gauge field~\cite{SF}. For $p_x+ip_y$ pairing the $\mu$ fermion forms a $U(1)_{4}$ topological order, corresponding $\nu_{Kitaev=2}$ in Kitaev's classification~\cite{Kitaev16}. Both bosons condense, $\langle d_I \rangle \neq 0$, and vortices of these condensates carry unit charges under fields $\beta_I$ which are dual to the $d_I$ currents. The gluing condition is that the neutral vortices must be accompanied by odd-strength $\pi$ vortices of both boson condensates. This leads to the topological superconductor action of Eq.~\eqref{Lsc} with the identification $\beta_\pm=\beta_1\pm\beta_2$.}:

\begin{equation}\label{Lsc2}
\begin{split}
&\mathcal{L}_{sc}=\frac{1}{\pi}\beta_0d\beta_0 +\frac{1}{2\pi}\beta_- d(a_1-a_2)+\frac{1}{2\pi}\beta_+d(a_1+a_2).
\end{split}
\end{equation}

\noindent where now a charge $l_0\in \mathbb{Z}\ {\rm mod}(4)$ under the gauge field $\beta_0$ labels the different quasiparticles of the fully neutral sector, so that the labels $\l_0=\{2,1,0,-1\}$ correspond to $\{\mu,v,1,\bar{v}\}$ respectively, which have the meaning of a complex fermion ($\mu$) and vortices with a zero mode filled or empty $(v,\bar{v})$ as discussed in Section~\ref{HLRpic}. On the other hand, a charge $l_{\pm}\in \mathbb{Z}$, under $\beta_{\pm}$ labels the two kind of vortices of the $\Delta$ and $\phi$ condensates, respectively, that trap flux $\pi l_{\pm}$ of $a^{\pm}_\mu$ respectively. Additionally, vortices of the neutral sector need to be glued to odd strength vortices of either of the charged sectors and the $\{1,\mu\}$ particles need to be glued to even strength vortices of the charged sectors. Namely, only quasiparticles in the sub-lattice $(l_0+l_-+l_+)/2\in \mathbb{Z}$ are physical. To implement this constraint we redefine gauge fields of the dual superconductor as: $\beta_0'=\beta_0+\beta_-$, $\beta_1'=\beta_++\beta_-$, $\beta_2'=-\beta_++\beta_-$, and enforce that the charges under these new gauge fields be integers. Upon integrating out the $a_{1,2}$ fields that glue the superconductor to the bosonic Laughlin sector one obtains $\beta_{1,2}'=\alpha_{1,2}$. The resulting $K$-matrix is 3$\times$3 and after a basis change implemented by:

\begin{equation}\label{W}
\begin{split}
&W=-\begin{pmatrix}
1 & 0 & 0\\
1 & 0 & 1\\
2 & 1 & 1
\end{pmatrix},
\end{split}
\end{equation}

\noindent with $W\in SL(3,\mathbb{Z})$, one obtains the following Chern-Simons theory:

\begin{equation}\label{KZ4}
\begin{split}
&\mathcal{L}=\frac{1}{4\pi}\alpha^T K d\alpha-\frac{1}{2\pi}A_+ t_+^T d\alpha-\frac{1}{2\pi}A_- t_-^T d\alpha+\cdots,\\
&K=\begin{pmatrix}
0 & 4 & 0\\
4 & 0 & 0\\
0 & 0 & -1
\end{pmatrix}, \ 
 t_+=
\begin{pmatrix}
2 \\
0 \\
-1
\end{pmatrix}, \ 
 t_-=
\begin{pmatrix}
0 \\
2 \\
-1
\end{pmatrix}.
\end{split}
\end{equation}

Therefore this state is fully gapped and has the topological order of $Z_4$ lattice gauge theory glued to a chiral integer quantum Hall state.  This state is exactly the same described in the previous section within the Dirac theory. In fact, the following is the correspondence between the labels of quasiparticles:

\begin{equation}\label{Qps}
\begin{split}
&V_+\leftrightarrow
\begin{pmatrix}
1\\
0 \\
0
\end{pmatrix}, \
\beta \leftrightarrow
\begin{pmatrix}
0\\
1 \\
0
\end{pmatrix}, \
c_1^\dagger \leftrightarrow
\begin{pmatrix}
0\\
0 \\
1
\end{pmatrix}.
\end{split}
\end{equation}

Additionally, within the HLR formulation it is possible to find out the transformation laws for the quasiparticle lattice under the layer exchange symmetry $X$, since this symmetry remains manifest. First we note that vorticity of the $\phi$ and $\Delta$ condensates are respectively odd and even under $X$, which follows from Eq.~\eqref{CTXphi} and the corresponding analogue of Eq.~\eqref{CTXDelta} for the HLR case. This implies that $X\beta_\pm X^{-1}= \pm \beta_\pm$. Additionally, the transformation property of the vortices of the neutral sector under $X$ dictates that their zero modes are filled or emptied upon its action, leading to: $X v X^{-1}\rightarrow \bar{v}$. This rule is implemented on the gauge fields as $X\beta_0 X^{-1}= - \beta_0$. Using this rules it is easy to find that the $X$ symmetry acts on the $K$-matrix of Eq.~\eqref{KZ4} as:

\begin{equation}\label{KZ4}
\begin{split}
&X: \alpha \rightarrow W_X \alpha, \ W_X^T K W_X=K, \ W_X=
\begin{pmatrix}
-1 & -2 & 1\\
 0 & -1 & 0\\
0 & -4 & 1
\end{pmatrix}.
\end{split}
\end{equation}

The rows of $W_X\in SL(3,\mathbb{Z})$ specify the transformation laws of the quasiparticles that serve as basis for the topological order listed in Eq.~\eqref{Qps}. It is reassuring to find that the transformation rules are exactly the same as those described within the Dirac picture in Eqs.~\eqref{CTXbeta} and~\eqref{CTXV}.
The discussion in this  and the preceding section illustrates that the $Z_4$ topological order with the anomalous particle-hole symmetry implementation that can be realized at the surface of a two-component chiral AIII topological insulator, as described in Ref.~\cite{Chong3D}, can also be realized in particle-hole symmetric two-component Landau levels.

It is interesting to note that if, starting from the layer decoupled limit, we had considered only dual composite fermion exciton condensation without pairing, namely $\langle \phi\rangle \neq 0$ and $\langle \Delta\rangle = 0$, we would induce an spontaneous composite fermion tunneling term that splits the two composite fermion fermi seas, which taking $\langle \phi\rangle \in \mathbb{R}$ would correspond to symmetric and anti-symmetric coherent superposition of composite fermions in the two layers. Therefore, this phase would correspond to the particle-hole symmetric version of the ICCFL state proposed in Ref.~\cite{Alicea09}. An important new qualitative feature that the Dirac nature of the composite fermion brings into this phase is that the tunneling term does not change the Berry phase of neither of the composite fermion surfaces, therefore both the symmetric and anti-symmetric composite fermion fermi surfaces would have a Berry phase of $\pi$ in such state. Figure~\ref{orders} summarizes the close relation between all these two-component particle-hole symmetric phases we have considered so far.

\section{Multicomponent particle-hole symmetric Landau levels}

In this section we will describe some interesting possible states in half-filled particle-hole symmetric Landau levels with four and eight components. 
Potential platforms with these many components are monolayer and bilayer graphene.  First note that just as in a single component system, a general $N$-component Landau level at half-filling can be fruitfully obtained in a {\em microscopic} system of $N$ massless Dirac fermions in a magnetic field. The Landau-level particle-hole symmetry is then obtained as an exact microscopic symmetry but the price to pay is that the microscopic system lives at the surface of a suitable three dimensional topological insulator. Since the physical situation is very similar to the single component case we will not elaborate on it here.  However there is one detail we will need to address.  We are interested in $N$-component Landau-level systems with (at least) $U(1) \times \cal{CT}$ symmetry.  Obtaining these through a microscopic Dirac theory then requires us to think about massless Dirac systems also with (at least) $U(1) \times \cal{CT}$ symmetry. For an $N$-component Dirac fermion with {\em only} this symmetry, it is known~\cite{Chong3D,Max3D} that there is no anomaly only if $N = 0 (\mod 8)$. Thus the Landau-level particle hole symmetry is anomalous for generic $N$ but not if $N = 8n$. For instance at $N = 8$ this means that the particle-hole symmetric Landau level can in principle be obtained in a strictly two dimensional microscopic model.  This situation changes once other symmetries are included as we discuss below.

 If the interaction Hamiltonian is just a density-density repulsion (such as Coulomb) then an $N$-component Landau level has $SU(N)$ symmetry.  The full symmetry of the half-filled $N$-component Landau level (including charge $U(1)$) is then $U(N)  \times \cal{CT}$.  With this higher symmetry we can again realize the Landau level in a microscopic system of Dirac fermions. Now we argue below that this Dirac fermion system is anomalous for all $N$ and not just when $N \neq  0 (\mod 8)$. Thus the physics of the $SU(N)$ symmetric $N$-component Landau level at any $N$ gets related to the physics of the surface of a three dimensional fermionic topological insulator with $U(N)\times \cal{CT}$ symmetry. 
 
 Much of the literature on such $N$-component Landau levels with $SU(N)$ symmetry  has focused on a quantum hall ferromagnets. While this is certainly a very natural state for the Coulomb Hamiltonian, it is interesting conceptually to consider other states that preserve some or all of the symmetries of the Hamiltonian.  
 We first prove that for $N$ even, a topologically ordered gapped state that preserves all the  symmetries is not possible. A symmetric gapped state may be  possible if the $SU(N)$ symmetry is either spontaneously broken or explicitly broken by the Hamiltonian to a smaller symmetry subgroup.\footnote{In the graphene examples residual terms that break the $SU(4)$ symmetry are always present but are small compared to the leading long range Coulomb interactions that respect $SU(4)$.} We illustrate this with some specific examples for $N = 4$ and $N = 8$.
 
 Our discussion of symmetric gapped states will use the perspective of quantum disordering a superfluid. In this approach one views a Mott insulating phase of interest as descending from a superconductor~\footnote{We view the system of interest as having an un-gauged probe electro-magnetic field, but we can gauge the probe fields as a technical device to facilitate the elucidation of the statistics of the several quasiparticles.} where the global $U(1)$ symmetry is restored via vortex condensation~\cite{SF}. This approach has been very fruitful for understanding symmetry protected topological order at the surface of topological insulators~\cite{SenthilReview}, and in particular it provides a simple route to understand the classification of the phases of the symmetry class AIII in the presence of strong interactions~\cite{Chong3D}. The discussion in this section follows closely that of Sect. V of Ref.~\cite{Chong3D} where the analysis assumed only a global $U(1)$ symmetry and the anti-unitary particle-hole $\mathcal{CT}$. Here we consider enlarged symmetries that are relevant to specific physical realizations.
 
 \subsection{$U(2N_f) \times \mathcal{C T}$ symmetry enforced gaplessness at $\nu=N_f$}\label{gapless}

Let us consider an even number of Landau levels, $2N_f$, at half filling, $\nu=N_f$, or, equivalently, the surface of an AIII topological insulator with $2N_f$ massless Dirac cones, and  restrict to the situation where there is  $U(2N_f) \times \mathcal{C T}$ symmetry.  We first argue that this is anomalous for any $N_f$. It suffices to show that the anomaly exists for massless Dirac fermions in zero background magnetic field (as the field does not change the symmetry).  For such Dirac fermions, consider the monopole operator\footnote{Strictly speaking we are weakly gauging the global $U(1)$ in thinking of the flux insertion as an operator.} associated with threading $2\pi$ flux of a background gauge-field $A$ that couples to the global $U(1)$ current. It is convenient to think of the Dirac theory as living at the interface between some $3d$ material and vacuum. Then the flux threading can be viewed as a process where a magnetic monopole from the outside vacuum tunnels into the material on the other side. The structure of this monopole operator in the Dirac theory  is well-known~\cite{borokhov}. For instance if the spatial surface on which the Dirac theory lives  is the surface of a sphere, a monopole configuration has $2N_f$ zero modes. Charge neutrality is achieved when $N_f$ of these are filled with fermions. It is easy to see that the monopole operator  is bosonic and transforms under the rank-$N_f$ fully antisymmetric representation of the $SU(2N_f)$ subgroup.  In particular  these operators transform non-trivially under the center $Z_{2N_f}$ of $SU(2N_f)$. On the other hand {\em local} operators that are charge neutral are built up as composites of the electron operator and will always transform trivially under this $Z_{2N_f}$ subgroup.  As usual the non-trivial transformation of the monopole insertion in the surface Dirac theory is allowed if the bulk 3d material  has the same non-trivial transformation for the bulk monopole.  It follows that the bulk is a non-trivial topological insulator for any $2N_f$.  Thus, as promised,  in the presence of additional global $SU(2N_f)$ symmetry the particle-hole symmetric  Landau level cannot be realized in any strictly $2d$ system.

We will now argue that it is impossible to construct a gapped phase that respects the full symmetry. The cornerstone of the argument is the observation that there exist no projective representations for $SU(2N_f)$. This implies that any topological order containing anyons ($x_I$) and the electron ($c$),  $\{1,x_I\} \times \{1,c\}$, is such that the anyons can always be taken to be $SU(2N_f)$ singlets. If an anyon has a non-trivial representation of $SU(2N_f)$ one can always replace it with an anyon bound to electrons such that the composite forms an $SU(2N_f)$ singlet without changing the symmetry realization and topological order. In addition, the action of $U(1)\times \mathcal{C T}$ must be closed within the topological sector. 

Moreover, if the phase realizes symmetry in an anomalous fashion, namely one that is not strictly allowed in a two-dimensional system with on-site symmetry implementations, then such an anomalous symmetry would have to be manifest at low energies in the topological sector $\{1,x_I\}$. Since the electric charge of any local $SU(2N_f)$ singlets is quantized in units $2N_f$, the minimal charge of a local operator constructed from fusing the anyons $\{1,x_I\}$ must be an integer multiple of $2N_f$. Moreover, since any local singlet is a bosonic operator, the topological order $\{1,x_I\}$ can be viewed as arising from a local bosonic singlet whose charge is some multiple of $2N_f$.

Let us now discuss what kind of excitation a fundamental monopole tunneling event would leave in such surface state. The bare electron, with charge $e$, experiences a magnetic flux quantum from a unit strength monopole in the bulk: $\Phi_e=\frac{h c}{e}$. A charge $2eN_f$ boson, will therefore experience an enlarged magnetic flux from the unit strength monopole in the bulk: $\Phi_b=2 N_f \frac{h c}{e}$. Therefore, the fundamental monopole of the bare electron is effectively $2N_f$-monopoles for the boson. For bosonic matter with $U(1) \times \mathcal{C T}$, the charge neutral monopoles with even strength are always trivial bosons ($\mathcal{C T}^2=1$) regardless of whether the bosonic bulk has a nontrivial $\theta=2\pi$ term~\cite{Max13}. The monopole would therefore be a charge neutral bosonic singlet transforming trivially under $\mathcal{C T} \times U(2N_f)$, and so would be an excitation at the surface created by the monopole tunneling event. However we saw above that the monopole transforms non-trivially under $Z_{2N_f}$. The assumption that the surface admits a symmetric gapped phase has thus produced a contradiction.

In Ref.~\cite{Chong3D} it was shown that for $N_f=4$, namely 8 Dirac cones, with only $U(1)\times \mathcal{C T}$ symmetry it is possible to construct a symmetric gapped state that has no topological order.  This implies that the gapless phase of 8 Dirac cones with only this symmetry can be  deformed through a phase transition at strong interactions into a trivial phase, equivalent to the surface of a $N_f = 0$ trivial bulk insulator. The argument given above shows that this is not possible in the presence of the larger $\mathcal{C T} \times U(2N_f)$ symmetry, and therefore that such symmetric $2N_f$ Dirac cones cannot be connected, without breaking the symmetry, to the trivial state.

\subsection{Four-components at $\nu=2$}\label{nu2}

Though with full $U(4) \times \mathcal{CT}$ symmetry, a symmetric gapped state is not allowed, we will show below that if the $SU(4)$ flavor symmetry is reduced to $SU(2) \times SU(2)$, preserving the overall $U(1) \times \mathcal{CT}$, then such a state is indeed possible. The four component particle-hole symmetric Landau level with these symmetries is still anomalous. Thus the proposed state - which strictly speaking cannot be realized in a $2d$ system- can nevertheless be realized (with the symmetries present with arbitrary precision) in the isolated Landau level.  We will then comment on the possible realization of such a state in monolayer graphene. 

\subsubsection{Maximally symmetric $eTmT$ state at $\nu=2$}\label{nu2}

 The state we discuss has the topological order of a $Z_2$ gauge theory but with an anomalous implementation of the $\mathcal{CT}$ symmetry~\cite{AshvinSenthil,ChongBosons,ChongScience,Chong3D}.  The $Z_2$ gauge theory has $3$ non-trivial quasiparticles $e, m, \epsilon$ which are all mutual semions. $e,m$ are bosons while $\epsilon$ is a fermion. With only $U(1) \times \mathcal{CT}$ symmetry the proposed  state  has $e$ and $m$ both transforming as Kramers doublets under $\mathcal{CT}$. For this reason it has been dubbed $eTmT$.  In the context of the present paper this state will be further `enriched' by the extra $SU(2) \times SU(2)$ symmetry.

 Our strategy for constructing this state is similar to previous papers~\cite{ChongScience,Chong3D}.  We will begin with a state with four massless Dirac fermions in zero magnetic field. We will break the global $U(1)$ symmetry by pairing them as follows:

\be\label{deltaH}
\delta H=i \Delta   \psi  \sigma_y \tau_y \mu_y \psi- i \Delta^* \psi^{\dagger}  \sigma_y  \tau_y \mu_y  \psi^{\dagger}.
\ee

where $\sigma$ and $\tau$ are Pauli matrices operating in the Dirac cone flavors. Even though the $U(1)$ symmetry is broken, this superconductor respects the combination $U(\pi/2) \ \mathcal{CT}$ and the $SU(2)_\tau\times SU(2)_\sigma$ symmetry of the separate rotations of the $\tau$ and $\sigma$ Pauli matrices, since the pairing is singlet with respect to either of those pseudospin flavors. 
We will then quantum disorder the superconductor by proliferating vortices, thereby restoring the broken $U(1)$ symmetry. The elementary $\pi$-vortex will have zero modes and will be non-trivial and cannot be proliferated while preserving the symmetry.  Quantum disordering the superconductor will require proliferating a higher strength vortex which will lead to a gapped topologically ordered state which inherits the anomalous symmetry of the original massless Dirac theory. As a non-zero magnetic field does not change the symmetry of the system, this topologically ordered state will also be a possible state of the half-filled four-component Landau level with the stated symmetries.

The zero modes in the vortex cores of this superconductor can be investigated in a similar spirit to the case of several Kitaev chains~\cite{Fidk}. The fundamental $\pi$ vortex contains four zero Majorana modes. We can combine these four Majoranas into two complex fermion zero modes. The local Hilbert space associated with filling these modes has dimension 4. Let us label these four states by the occupation numbers of these complex modes $|n_1,n_2\rangle$, where $n_{1,2}=\{0,1\}$. One can choose the complex zero modes such that the subspace with a singly occupied mode, $\{|1,0\rangle, |0,1\rangle\}$, transforms as a spin $1/2$ representation under the $SU(2)_\tau$ transformations while transforming as trivial singlets under the $SU(2)_\sigma$. Then one finds that the complementary subspace, $\{|0,0\rangle, |1,1\rangle\}$, would form a spin $1/2$ representation of $SU(2)_\sigma$ while transforming as trivial singlets under $SU(2)_\tau$. In other words, one can show that the zero mode Hilbert space decomposes into a $(1/2,0)\oplus(0,1/2)$ representation of $SU(2)_\tau\times SU(2)_\sigma$. This implies that the $\pi$ vortex is forced to carry non-trivial quantum numbers of these symmetries.

However, by combining two of these $\pi$ vortices one can construct a state that is an $SU(2)_\tau\times SU(2)_\sigma$ singlet and hence transforms trivially under all the symmetries that remain present in the superconductor state. As a consequence such $2\pi$ vortex would behave as a trivial boson which can be condensed to restore the $U(1)$ symmetry, and consequently the $\mathcal{CT}$, resulting in an insulating phase with the topological order of $Z_2$ gauge theory~\cite{SF} enriched by a large symmetry: $U(1) \times \mathcal{CT} \times SU(2)_\tau\times SU(2)_\sigma$.\footnote{Formally the symmetry group is the one written above mod $\mathbb{Z}_2^2$, which avoids double counting of the $(-1)$ elements.}

Several non-trivial deconfined quasiparticles are present in this insulator. There will be a neutral fermion (spinon) which is the remnant of the Bogoliubov fermion and we label $\epsilon$, a boson (chargon) labeled $h$, and two remnants of the $\pi$ vortex (visons) labeled $\{m,e\}$. The visons $\{m,e\}$ are bosons. $m$ can be chosen as descending from the vortex states in which all the complex zero modes are half-filled $\{|1,0\rangle, |0,1\rangle\}$ and hence it carries pseudo-spins $s_\tau=1/2$ and $s_\sigma=0$. $e$ can be chosen as descending from the vortex states $\{|0,0\rangle, |1,1\rangle\}$ and hence it carries pseudo-spins $s_\tau=0$ and $s_\sigma=1/2$. It follows that the spinon, $\epsilon=e\times m$, carries a fundamental representation of $SU(2)_\tau\times SU(2)_\sigma$ with $s_\tau=1/2$ and $s_\sigma=1/2$. The chargon, $h$, will be a trivial object under these symmetries as it is essentially a descendant of half a cooper pair and the cooper pair field is a singlet under these symmetries. However, $h$ will carry physical charge $N_+=1$ (same as the physical electron) under the restored $U(1)$ charge conservation. The physical electron is therefore $c =h \times \epsilon$.   This state has also been discussed in Ref.~\cite{Nahum} in a different context.

This state implements the $U(1)\times \mathcal{CT}$ in an anomalous fashion in which the $e$ and $m$ particles are charge neutral Kramers bosons $\mathcal{CT}^2=-1$ as can be seen following similar arguments to Ref.~\cite{Chong3D}. The state in question is therefore an $SU(2)\times SU(2)$ invariant version of the $eTmT$ state previously considered in the literature~\cite{AshvinSenthil,SenthilReview,ChongBosons,ChongScience,Chong3D}. Additionally, this state has an interesting discrete symmetry that exchanges the $e$ and the $m$ particles. Consider the following symmetry operation that exchanges the $\tau$ and $\mu$ pseudo-spin flavors:
\begin{equation}\label{Lambda}
\begin{split}
& \psi \rightarrow \Lambda \psi, \ \Lambda \equiv\frac{1}{2}\sum_{\nu=0}^{3}\tau_\nu \sigma_\nu,\\
& \Lambda=\Lambda^T=\Lambda^\dagger,\ \Lambda^2=1, \ \Lambda\tau_\mu \sigma_\nu \Lambda= \tau_\nu \sigma_\mu.\\
\end{split}
\end{equation}
\noindent The last property of $\Lambda$ implies that the pairing $\delta H$ from Eq.~\eqref{deltaH} respects this symmetry. This symmetry acts by exchanging the $\tau$ and $\sigma$ quantum numbers, and, hence, it exchanges the $e$ and $m$ particles. Even though we know of no specific potentially realistic physical system possesssing all the symmetries we considered here, this phase is a good starting point from which lower symmetry incarnations of the $eTmT$ phase can be conveniently understood.
 
\subsubsection{Monolayer graphene}

The zeroth Landau level of graphene is four-fold degenerate and the problem of interacting electrons projected onto this Landau level can be viewed at low energies as a theory of the surface of AIII topological insulator with four Dirac cones in the strong magnetic field limit, as the system we just described. A good model Hamiltonian for graphene in this limit includes the long-ranged Coulomb interaction, two types of short-ranged interactions that account for lattice scale interactions~\cite{Kharitonov,Fengcheng}:

\begin{equation}\label{gzgperp}
\begin{split}
&V^{coul}_{i j}=\frac{e^2}{\epsilon \left|r_i-r_j\right|},\\
& V^{latt}_{i j}=(g_z \tau _i^z \tau _j^z +g_\perp (\tau _i^x \tau _j^x+\tau _i^y \tau _j^y))\delta^{(2)}(r_i-r_j),
\end{split}
\end{equation}

\noindent and the Zeeman coupling. Here $\tau$ denote Pauli matrices in valley space, and $g_z$ and $g_\perp$ are parameters characterizing the strength of valley-dependent interactions. From these terms the Coulomb interaction is by far the most dominant. The projected Hamiltonian with only Coulomb interaction has $SU(4)$ symmetry in addition to $U(1) \times \mathcal{C T}$. It is believed that the short-ranged interactions are typically stronger than the Zeeman term by roughly an order of magnitude~\cite{Kharitonov,Andrea12,Andrea14,Abanin,Inti,Fengcheng}. These terms break the $SU(4)$ symmetry into $SU(2)_{spin}\times (U(1)\rtimes X)_{valley}$, where $X$ here denotes a discrete $\mathbb{Z}_2$ valley exchange symmetry analogous to the layer exchange considered in Section~\ref{2lyrint}. Importantly, these interactions preserve the anti-unitary particle-hole symmetry $\mathcal{C T}$. On the other hand, the Zeeman term breaks the internal spin-valley symmetries further down to $U(1)_{spin}\times (U(1)\rtimes \mathbb{Z}_2)_{valley}$, and, more crucially, it destroys the anti-unitary particle-hole symmetry $\mathcal{C T}$.

Therefore in order to view the interacting Hamiltonian of graphene as a special limit of a topological insulator surface one needs to neglect the Zeeman term. In this context it is possible that the $eTmT$ symmetry enriched topological order arises in graphene at neutrality. This state will be a version of that described in Sec.~\ref{nu2} with its symmetry properly reduced to $U(1) \times \mathcal{C T}\times SU(2)_{spin}\times (U(1)\rtimes X)_{valley}$. We note that a weak breaking of $\mathcal{C T}$ symmetry, such as that expected from Zeeman, will split the excited states whose degeneracy relies on their Kramers nature under $\mathcal{C T}$, however since $eTmT$ is a gapped phase its ground state will be only weakly modified under small $\mathcal{C T}$ breaking terms.

There exist strong numerical evidence supporting that the ground states of the projected Coulomb plus short-ranged interactions Hamiltonian are quantum Hall ferromagnets~\cite{Fengcheng}. Experiments have found that upon increasing the Zeeman term via in-plane magnetic fields a relatively smooth transition, during which the bulk charge gap remains open, into a state consistent with a ferromagnetic order occurs~\cite{Andrea14}. From the candidate quantum Hall ferromagnets the one that appears most consistent with this picture is the anti-ferromagnet. However, considering the fact that Landau level mixing is expected to be strong in graphene~\cite{Peterson} and its effects on the energetics of quantum Hall ferromagnets have not been well explored, it appears reasonable not to rule out the possibility that they could stabilize exotic states such as the $eTmT$. Experimentally an $eTmT$ state would look like a trivial integer quantum Hall state from the point of view of charge transport, but it would be non-trivial in the neutral sectors. This makes challenging detecting the $eTmT$ state in graphene, but also ruling it out on experimental grounds. The transition from the $eTmT$ into conventional quantum Hall ferromagnets would be driven by the condensation of one non-trivial bosons. For example, upon increasing the Zeeman term one expects that the boson carrying the spin $1/2$, e.g. the $e$ particle, would condense and hence drive a confinement transition for the $m$, $\epsilon$ and $h$ particles, while breaking the spin rotation symmetry resulting in a trivial integer quantum Hall state with ferromagnetic ordering.

\subsection{Eight-components at $\nu=4$}\label{nu4} 
We now briefly consider $8$-component Landau levels which at half-filling do not have anomalous implementation of $U(1) \times \mathcal{CT}$ but have anomalous implementation of $U(8) \times \mathcal{CT}$. We will 
study the possibility that the anomaly disappears for some subgroup of $U(8)$ that is bigger than just the $U(1)$.

\subsubsection{Highly symmetric gapped state with no topological order}\label{nu4}

  We will show that if the symmetry is $U(1)\times \mathcal{CT}$ but with $SU(8)$ reduced to $SU(2)\times SU(2)\times O(2)$ then a symmetric gapped state with no topological order is possible. This implies that the $8$-component Landau level with these symmetries is not anomalous and hence can be obtained microscopically in a strictly $2d$ system. 
Following the strategy of the previous section we  start with $8$ massless Dirac fermions and consider a superconductor described by the following pairing~\cite{Chong3D}:

\be\label{deltaH}
\delta H=i \Delta   \psi  \sigma_y \tau_y \mu_y \nu_0 \psi- i \Delta^* \psi^{\dagger}  \sigma_y  \tau_y \mu_y \nu_0 \psi^{\dagger}.
\ee

\noindent 
where $\nu$ are Pauli matrices in an additional pseudo-spin flavor. Notice that this pairing term is not compatible with a full $SU(2)$ symmetry on the $\nu$ pseudo-spin, and it is important that the flavor symmetry on the $\nu$ index is only $O(2)$~\footnote{This remaining symmetry can be viewed as the $U(1)$ subgroup generated by $\psi^\dagger \rightarrow e^{i \frac{\theta}{2} \nu_y} \psi^\dagger$ and the $\mathbb{Z}_2$ symmetry of exchange of $\nu$ flavors $\psi^\dagger \rightarrow \nu_x \psi^\dagger$.}. It is easy to see that the fundamental $\pi$ vortex can be taken to have a trivial gapped core. Thus this vortex can be condensed and we obtain the promised symmetric gapped state without topological order.  This state preserves a large symmetry group, including, notably, the anti-unitary particle-hole symmetry $\mathcal{CT}$, and is a possible state in an $8$-component Landau level with this symmetry. Notice that any fermion bilinear that selects a unique integer quantum Hall state would necessarily break the particle-hole symmetry as it would gap the surface of the topological insulator, and in this sense this state cannot be described by a simple mean-field Hartree-Fock state. In addition to this continuous symmetries we would also have the discrete set of permutations between the two $SU(2)$ flavors, in analogy to the $\Lambda$ symmetry featured in Eq.~\eqref{Lambda}. Since the state in question is a fully gapped insulator, explicit terms in the Hamiltonian which weakly break any of the symmetries are expected to lead only to small adiabatic changes of the ground state.

\subsubsection{Bilayer graphene}

AB-stacked bilayer graphene has a special electronic dispersion which renders its zero Landau level eight-fold degenerate~\cite{Novoselov,Falko}. In addition to spin-valley degeneracy, this zero Landau level contains degenerate cyclotron orbitals $n=0$ and $n=1$. Because these orbitals have different form factors even the projected Coulomb interaction into the zero Landau level has no symmetry operations rotating between the $n=0$ and $n=1$ orbitals and simply has an $SU(4)$ spin-valley symmetry.  

A more subtle issue is the particle-hole symmetry in the zero Landau level of bilayer graphene~\footnote{Particle-hole symmetry is understood to map states at filling $\nu$ to states at $-\nu$, where $\nu$ is the filling measured from neutrality.}. Experiments have found particle-hole asymmetric sequences of fractional quantum Hall states~\cite{Yacoby}. Several authors have incorrectly assumed that the symmetry is broken by the Coulomb interaction itself projected into the zero Landau level, because of the different form factors of the degenerate $n=0$ and $n=1$ orbitals. In fact, as pointed out in Refs.~\cite{Shizuya}, there is a non-trivial Coulomb interactions with the negative energy sea of occupied states in bilayer graphene that is needed to properly account for the particle-hole symmetry~\cite{Rohit,thesis}. A likely explanation behind the particle-hole asymmetry observed in the experiments of Ref.~\cite{Yacoby} are intrinsic and sample-specific terms that break the particle hole symmetry~\cite{thesis}.

The special kind of particle-hole symmetry that we need in order to view the zero Landau level of bilayer graphene as the surface of an AIII topological insulator with eight Dirac cones is still even more restrictive. In addition to neglecting these terms that break the lattice particle-hole symmetry we need to neglect any single particle term that tends to select a trivial integer quantum Hall state at neutrality. Therefore, just as in the case of monolayer graphene, we need to neglect the Zeeman term and the interlayer bias. The interlayer bias is an experimentally tunable parameter, so, it can always be tuned to zero, while neglecting the Zeeman term is an approximation. 

The valley-dependent lattice scale interactions will also be present in bilayer graphene and assuming they have zero range they will have the same form as those in Eq.~\eqref{gzgperp} describing monolayer graphene~\cite{Kharitonovbilayer}. In this limit, the symmetry of the Hamiltonian of neutral bilayer graphene would be $U(1) \times \mathcal{C T}\times SU(2)_{spin}\times (U(1)\rtimes X)_{valley}$, just as in the monolayer. In order to realize the $\mathcal{C T}$-symmetric state described in Sec.~\ref{nu4}, we additionally need that the $\mathcal{C T}$ symmetry is not broken spontaneously. To our knowledge, there is no exact diagonalization or density-matrix-renormalization-group study of the full 8-fold degenerate zero Landau level including explicitly the $n=0$ and $n=1$ orbitals that would explore in an unbiased manner which type of ground state the Coulomb interactions would choose. Experimentally there is clear evidence for a gapped ground state at neutrality in bilayer graphene~\cite{Weitz,Maher}, and this state is consistent again with an anti-ferromagnetic quantum Hall ferromagnet state~\cite{Kharitonovbilayer}. But again we would like to emphasize that given the lack of complete numerical studies in bilayer graphene it is not ruled out that this state could be a descendant of the particle-hole invariant highly symmetric phase described in Sec.~\ref{nu4}.

\section{Summary and Discussion}
We have shown that the familiar exciton condensate experimentally realized in GaAs quantum Hall bilayers can be alternatively viewed as an interlayer paired state of composite fermions in a special channel that preserves particle-hole symmetry. This identification is a new application  of the fermionic particle-vortex duality. The quantum Hall bilayer at $\nu=1/2+1/2$ is an insulator with respect to the symmetric layer charge but a superfluid with respect to the layer charge imbalance. We showed that alternately it can be viewed as a ``superconductor"  with respect to the symmetric composite fermion density but an insulator with respect to the composite fermion layer density imbalance, {\em i.e.} as an interlayer paired composite fermion state. We showed that such a dual description of the phase can be understood either from the Dirac or HLR pictures, although only the former allows for the particle-hole symmetry   to be manifest. Further, we showed that elementary meron vortices of the exciton condensate serve as a basis out of which all other gapped quasiparticles can be obtained as bound states. Out of these defects there exists a $4\pi$ vortex that is charge neutral and has a Kramers structure under the anti-unitary particle-hole symmetry that survives in the exciton condensate, denoted $\tilde{\mathcal{CT}}$. This particle is the closest incarnation of the composite fermion itself, since it is simply the Bogoliubov fermion resulting from the interlayer composite fermion pairing. In this sense the exciton condensate offers us a rather unexpected window into the physics of the half-filled Landau level itself.

Determining the exact ground state of the ideal quantum Hall bilayer (the problem of Coulomb interacting electrons projected to the lowest Landau level with negligible interlayer tunneling) is a difficult problem at arbitrary interlayer distances. However, numerical studies suggest that the ground state at intermediate distance can be described by a paired state which has precisely the pairing channel considered here~\cite{Moller,Milo}. As has been previously pointed out~\cite{Bonesteel}, and as we have argued employing an RG analysis, there exist a weak coupling instability to interlayer Cooper pairing in the limit of infinite layer separation. This suggests the natural conjecture that perhaps the ground state of the ideal quantum Hall bilayer never encounters a quantum phase transition as a function of interlayer distance and has a smooth crossover from a BEC-like limit at small distances to a BCS-like limit at larger inter-layer distance. However as mentioned in the Introduction, a very recent Eliashberg calculation of the pairing symmetry in the large-$d$ limit~\cite{isobefu} finds a pairing channel different from the previous numerical work. Further numerical studies of realistic quantum hall bilayers is clearly called for.

We also describe a potential alternative ground state for a quantum Hall bilayer which can be thought of as a quantum disordered version of the exciton condensate. This state is fully gapped and preserves all the microscopic symmetries. It is likely the minimal state with these properties and has the topological order of a $Z_4$ gauge theory with an anomalous implementation of particle-hole symmetry. In this state the merons are liberated from their logarithmic energy cost and become fully deconfined quasiparticles. This exotic phase additonally features the presence of a fractional exciton quasiparticle which is essentially a quarter of the familiar interlayer electron-hole pair. It is for future studies to determine if suitable perturbations could realistically bring such a phase into experimental realization.

Finally we studied some aspects of  half-filled Landau levels of systems with $N = 4$ or $N = 8$ component fermions.  If these Landau levels have full $SU(N)$ symmetry (so that the full symmetry including particle-hole is $U(N) \times \mathcal{CT}$), then the symmetry realization is anomalous. We showed the impossibility of symmetry preserving gapped ground states in such a system. Thus if the symmetries are preserved, then the ground state must be a gapless liquid. Alternately the symmetry may be spontaneously broken as in the familiar quantum Hall ferromagnet. If the microscopic symmetry is smaller, then a symmetry preserving gapped state may be possible. We illustrated this with some examples for $N = 4$ and $N = 8$.  We did not however attempt to understand the microscopic situations that will facilitate the appearence of such states, and this is an interesting target for future work.

{\it Note added}: For complementary work on multicomponent half-filled quantum Hall systems, developed in parallel to ours by Potter, Wang, Metlitski and Vishwanath, see Ref.~\cite{Drewpaper}.


\section{Acknowledgements}

We thank Liujun Zou, Hiroki Isobe, Zheng Zhu, Liang Fu, Zlatko Papi\'c, Maissam Barkeshli and Eduardo Fradkin for very helpful discussions. We also thank Hiroki Isobe and Liang Fu and Andrew Potter, Chong Wang, Max Metlitski and Ashvin Vishwanath for sharing their results prior to publication.  I.S.\ and I.K.\  are supported by the Pappalardo Fellowship at MIT, C.W.\  is supported by the Harvard Society of Fellows, and T.S.\   is supported  by a  US Department of Energy 
grant DE-SC0008739. T.S.\  was also partially supported by a Simons Investigator award from the Simons Foundation. 

\appendix

\section{Dual vortices of the electron exciton condensate in Dirac picture}\label{dualBdG}

The mean-field BdG Hamiltonian for the vortices in the neutral sector of the superconductor of composite fermions described in Section~\ref{Dirac} reads as:

\be
H_{BdG}=\psi^\dagger (p_x\sigma_x+p_y\sigma_z-\mu)\psi + \frac{\Delta(r)^*}{2} i \psi^\dagger \tau_x\sigma_y \psi^\dagger+{\rm h.c.},
\ee

\noindent where $\psi$ carries indices $1,2$ denoting layer in addition to Dirac pseudo-spin indices. Let us define a BdG destruction operator as follows:

\be
\varphi=\begin{pmatrix}
\psi_1 \\
i \sigma_y \psi_2^\dagger
\end{pmatrix},
\ee

\noindent allowing to write the BdG equation as follows:

\be
H_{BdG}=\varphi^\dagger 
\begin{pmatrix}
p\cdot \sigma -\mu& \Delta(r)^* \\
\Delta(r) & -p\cdot \sigma+\mu 
\end{pmatrix} \varphi.
\ee

\noindent This BdG equation is formally identical to that of the Fu-Kane superconductor~\cite{FuKane}, however, we have not ``doubled-counted" particles and holes since $\varphi$ destroys particles in layer $1$ and holes in layer $2$. As a consequence every eigen-mode of the BdG problem (with positive, negative or zero energy) can be interpreted as a conventional complex fermion mode (two Majorana modes). In a vortex of vorticity $n$ the paring field has the form $\Delta({\bf r})=\Delta(r) e^{in\theta}$. It follows that the odd-strength vortices of $\Delta$ have one complex zero mode.

Notice that $\varphi^\dagger$ carries a definite charge of $-1$ under the  $a^{-}_\mu$ gauge field. Therefore all the vortex states can be uniquely labeled with $a^{-}_\mu$ charge. Additionally the BdG Hamiltonian is invariant under the layer exchange symmetry: 

\be
X \psi X^{-1}=\tau_x \psi, \ X \varphi X^{-1}=\varphi^\dagger \tau_x i \sigma_y. 
\ee

\noindent The $a^{-}_\mu$ charge, $\hat{q}_-\equiv \int d^2 {\bf r}\  \psi^\dagger \tau_z \psi$, is odd under $X$: $X\hat{q}_- X^{-1}=-\hat{q}_-$. Because $X$ acts as a particle-hole on $\varphi$, we conclude that the fundamental vortex with the complex zero mode empty, $V_-$, must have a charge $q_-=1/2$, whereas the vortex with the zero mode filled, $V_+$, must have a charge $q_-=-1/2$.

\section{Microscopic wavefunctions for exciton order parameter vortices}\label{Evortices}

Vortices of the exciton condensate are well studied in the quantum Hall literature. One approach is to start from the $SO(3)$ symmetric $\nu=1$ quantum Hall ferromagnet described by a non-linear sigma model and consider its XY limit~\cite{Moon}. In this model one can infer the fractional charge of the merons (XY vortices) starting from the relation~\cite{Sondhi}:

\be
N_+=-\frac{1}{4 \pi }\int d^2{\bf r} \ \hat{{\bf t}}\cdot \left(\frac{\partial \hat{{\bf t}}}{\partial x} \times \frac{\partial \hat{{\bf t}}}{\partial y}\right),
\ee

\noindent where $\hat{{\bf t}}$ is the unit-vector order parameter of the ferromagnet. For a vortex with an order parameter winding of $2 \pi w$ ($w \in \mathbb{Z}$), since $\hat{t}_z(\infty)=0$, one gets that it carries a half-integer quantized charge $N_+=-w \hat{t}_z(0)/2$, where $\hat{t}_z(0)=\pm1$ is the orientation of the order parameter at the vortex core.

In this section we will provide alternative explicit microscopic description for the exciton condensate order parameter vortices that allows to understand various properties in a straightforward way. These wavefunctions can be thought as the ones corresponding to the limit of smallest possible vortex cores and are perhaps not energetically favorable when the layer spacing is much smaller than the magnetic length, but might be favorable when the layers are farther apart so that the Coulomb capacitive energy penalizes severely the deviations of the order parameter away from the XY plane shrinking the vortex cores to small sizes. We emphasize, however, that our primary interest concerning these wavefunctions is not their energetics but rather their conceptual simplicity for illustrating various universal properties.

\begin{figure}[t]
	\begin{center}
		\includegraphics[width=3.5in]{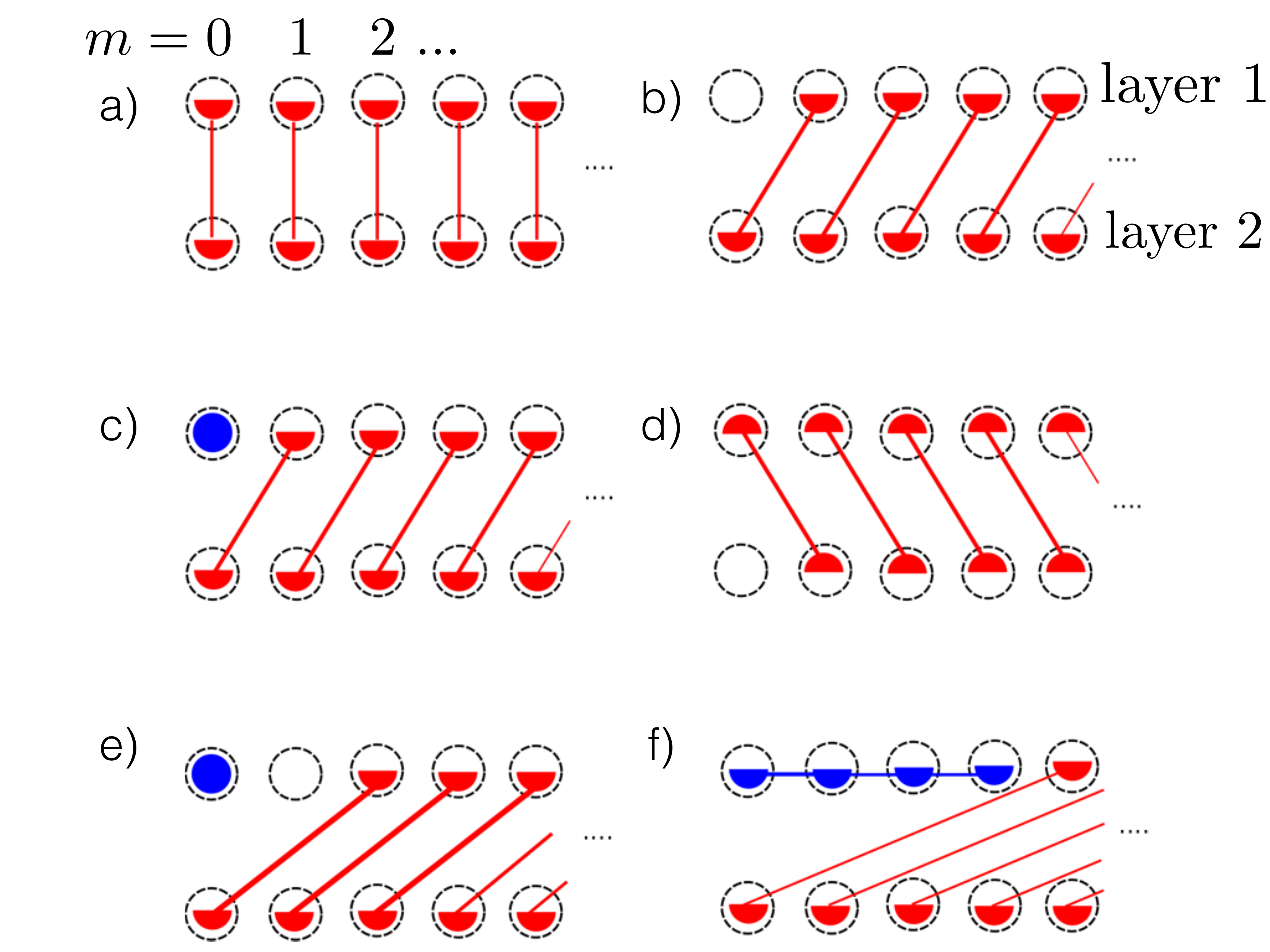}
	\end{center}
	\caption{(Color Online) Schematic representation of the many-body vortex states. The dashed circles are electron states with orbital $m$ and layer index $1,2$. Figure a) is the exciton ground state where electrons occupy states with definite orbital in a superposition of both layers with equal amplitude, which is represented by the half-circles joined by straight line. b) Meron vortex with charge $N_+=-1/2$ and vorticity $w=1$. c) Meron vortex with charge $N_+=1/2$ and vorticity $w=1$, where the zero mode of b) is occupied by an electron depicted as a blue ball. b) Meron vortex with charge $N_+=-1/2$ and vorticity $w=-1$, obtained from a layer swap, $X$-operation, from b). e) The charge neutral $N_+=0$ and $w=2$ vortex, which is obtained by filling one of the zero modes and is a Kramers neutral fermion and the closest incarnation of the Dirac composite in the exciton condensate. f) A particle-hole invariant charge neutral $w=4$ vortex corresponding the composite fermion particle-hole pair (exciton) whose condensation, considered in Section~\ref{Z4}, drives the transition into the $Z_4$ ordered state. This vortex is obtained by filling the zero modes with two fermions in a unique state analogous to the filling of 8 Majorana chains ends by Fidkowski and Kitaev~\cite{Fidk}.}
	\label{vortices}
\end{figure}

We begin by writing a mean-field single-particle Hamiltonian for the electron exciton condensate projected to the Lowest Landau level:

\be
H_{MF}=P_{0}(t_x({\bf r}) \tau _x+t_y({\bf r}) \tau _y)P_{0},
\ee

\noindent where $\tau$ are Pauli matrices acting in the layer index and $P_0$ is a projector into the lowest Landau level. The Hamiltonian contains no kinetic energy but only coupling to the spatially dependent XY order paramter $t_{x,y}({\bf r})$. Consider now the configuration for a circularly symmetric vortex centered at the origin: $t_x({\bf r}) +i t_y({\bf r})=t(r) e^{i w \theta}$, with $(r,\theta)$ polar coordinates for ${\bf r}$, and $w \in \mathbb{Z}$. Recall that in the symmetric gauge the single particle wavefunctions in the lowest Landau level take the form:

\be
\phi_m(r,\theta)=\frac{r^m e^{im\theta}}{\sqrt{2^{m+1} \pi  m!} l^{m+1}}e^{-\frac{r^2}{4 l^2}}, \ m=0,1,2,...
\ee

\noindent For $w>0$ it is then easy to verify that the following are non-zero energy eigenstates of the mean-field Hamiltonian:

\be
\psi_{m s}(r,\theta)=\frac{1}{\sqrt{2}}\left(
\begin{array}{c}
 \phi _m(r,\theta) \\
 s \phi _{m+w}(r,\theta) \\
\end{array}
\right), \ s=\pm 1, \ m=0,1,2,...
\ee

\noindent where the components of the column vector correspond to top and bottom layers. The mean-field energy of these modes is:

\be
E_{ms}=s \int_0^{\infty } \frac{e^{-x}  x^{m+\frac{w}{2} } t(r=l \sqrt{2 x})}{2 \sqrt{m! (m+w)!}} \, dx.
\ee

\noindent where $E_{m,-}<0$. In addition there are $w$ complex fermion zero energy modes:

\be
\psi_{m 0}(r,\theta)=\left(
\begin{array}{c}
 0 \\
 \phi _{m+w}(r,\theta) \\
\end{array}
\right), \ m=-w,...,-1.
\ee

\noindent A similar structure is found for $w<0$, in which case the zero modes are localized in the top layer. Now, if we construct the many-body vortex state by filling all the negative energy eigenstates ($s=-1$, $m\in \mathbb{Z}$), it is easy to verify that the vortex core has a deficit of $w/2$ particles relative to the ground state with no vortices (which corresponds to $w=0$), and therefore carries charge $N_+=-w/2$. Figure~\ref{vortices} illustrating  the many-body vortex state makes this transparent.

Various properties can be explicitly understood in terms of these vortices. For example, the layer exchange symmetry changes the vortex texture as: 

\be
X:\  t_x+it_y\rightarrow t_x-it_y,
\ee

\noindent therefore it leaves the physical charge $N_+$ invariant, but changes the vorticity $w\rightarrow -w$. This is the same statement of the fact that the $a^{-}_\mu$ charge $q_-$ is odd under layer exchange as described in Appendix~\ref{dualBdG} in the dual picture. Also the odd-$n$ strength vortices are forced to carry physical charge $N_+$.

The closest incarnation of the composite fermion in the dual picture is the bogoliubov fermion $\epsilon$. This object is charge neutral, $N_+=0$, but carries vorticity of the exciton order parameter. As described in Section~\ref{Dirac} we expect it to be a vortex with $4\pi$ winding of the order parameter, hence we choose $w=2$. Such vortex has two-complex fermion zero modes. When the two zero modes are empty this object carries charge $N_+=-1$. Therefore the composite fermion vortex is obtained by filling one of these two zero modes. The microscopic particle-hole symmetry acts on the vortex texture as:

\be
\mathcal{CT}:\  t_x+it_y\rightarrow -(t_x+it_y),
\ee

\noindent This symmetry is broken in the ground state as it reverses the magnetization. However the closely related operation $\tilde{\mathcal{CT}} = \mathcal{CT} U_{1}(\frac{\pi}{2})U_2(- \frac{\pi}{2})$ remains a particle-hole symmetry. We can choose these symmetries to act on the electron operators as:

\begin{equation}\label{Kmat}
\begin{split}
&\mathcal{CT} c_{a m}\mathcal{CT}^{-1}=i c^{\dagger }_{a m},\\
&U(\phi)_z c_{a m} U(\phi)_z^{-1}=(e^{-i \phi \tau_z})_{ab} c^{\dagger }_{b m},
\end{split}
\end{equation}

\noindent where $c_{am}^\dagger$ creates an electron in layer $a=\{1,2\}$ and orbital $\phi_m$. One finds then that the action of $\tilde{\mathcal{CT}}$ on the zero modes to be:

\be
\tilde{\mathcal{CT}} \psi_{m 0} \tilde{\mathcal{CT}}^{-1}=-\psi_{m 0}^\dagger
\ee

\noindent This symmetry squares to $\tilde{\mathcal{CT}}^2=1$ acting on electron operators. If we denote $|p_1,p_2\rangle$ the many-body state corresponding to the $4\pi$ vortex of the order parameter, with $p_i=\{0,1\}$ denoting the occupation of the zero modes, we will have that there are two charge neutral states, namely $\{|1,0\rangle,|0,1\rangle\}$. These two states are mapped into one another by $\tilde{\mathcal{CT}}|1,0\rangle=|0,1\rangle$, moreover, from the action of $\tilde{\mathcal{CT}}$ on the zero modes one concludes that it squares to $\tilde{\mathcal{CT}}=-1$ on the vortex states $\{|1,0\rangle,|0,1\rangle\}$. Therefore this symmetry has a projective representation on these vortices. This is the manifestation of the Kramers structure of the composite fermion.

\section{Particle-hole symmetry in the exciton condensate: alternate view}
\label{altCT}
In this Appendix we show how the $\tCT$ properties of the exciton condensate can be obtained in an alternate point of view  through a construction directly in terms of electrons. For non-relativistic electrons, $\CT$ and $\tCT$ are symmetries only when the Hamiltonian is projected to the lowest Landau level. The associated large degeneracy of single particle states makes an analysis difficult. Here we will follow a different approach analagous to that used in recent discussions of particle-hole symmetry in single component systems. We will take our microscopic electron system to be two flavors of massless Dirac electrons with $\CT$ symmetry.  This is realized as the surface state of a $3d$ chiral topological insulator (in class AIII but with an additional $U(1)$ symmetry corresponding to separate conservation of both flavors of electrons). Specifically, the Lagrangian is 
\be
{\cal L} = \sum_I \bar{\chi}_I i \slashed{\cal D}_A \chi_I + {\cal L}_{int}
\ee
Here $\chi_I$ are each $2$-component Dirac electrons, and $I = 1,2$ is the flavor index. $A$ is a background gauge field, and $\slashed{\cal D}_A $ is the Dirac operator. This is $\CT$ invariant if we let $\chi_I \rightarrow i\gamma_0 \chi^\dagger_I$, and change $A_0 \rightarrow - A_0, A_i \rightarrow  A_i$.  

A non-zero magnetic field $B$ does not break any symmetries, and hence can be included.  There will be two zero energy Landau levels which will each be half-filled due to the $\CT$ symmetry. Projecting to these levels, we get the $\nu = 1/2 + 1/2$ quantum Hall bilayer with $\CT$ symmetry that we are interested in. 

Here we will study the exciton condensate phase in this system in {\em zero} $B$-field.  We will make the reasonable assumption that this $B = 0$ exciton condensate is smoothly connected to the one that obtains in the large-$B$ limit. Indeed, we will see that the excitation structure and symmetry properties are identical to that in our earlier constructions. 

The $B = 0$ exciton condensate we study will have a gap to all fermion excitations.  We characterize it by an order parameter $\sim e^{i\theta}$. As usual this breaks $\CT$ but preserves $\tCT$.  The most obvious excitation is the relic of the $\chi$ fermion which is gapped.  We strip off it's $N_-$ charge, and call the resulting fermion $f$. This will have $N_+ = 1$.  The condensate will also have vortex excitations associated with $2\pi w$ winding of $\theta$, $w\in \mathbb{Z}$.  The $f$ particle will have mutual $\pi$ statistics around all odd $w$ vortices, and will be local around even $w$ vortices. 

A $2\pi$ vortex in $\theta$ is readily seen to have a single complex $0$ mode. Thus there are two such vortices that differ by the addition of $f$.  We call them $V_+$ and  $\bar{V}_-$ (as we will shortly identify them with objects denoted by the same symbols in the dual construction described in the main text).  Note that their $N_+$ charges must differ by $1$. Further $\tCT$ interchanges these two vortices. Thus these vortices must have $N_+ = \pm 1/2$. 
Note also that they are mutual semions as they differ by the binding of $f$ which is a mutual semion around either of them. These are exactly the right properties of the $V_+$ and  $\bar{V}_-$  as described in the construction of Section~\ref{Dirac}. 

Next consider $4\pi$ vortices. These harbor two complex zero modes, and can be analysed by studying their various possible fillings. It is simpler however to obtain them as composites of the $2\pi$ vortex. The logic is now completely similar to our earlier construction, and we will get an electrically neutral $4\pi$ vortex $\epsilon$ that is Kramers under $\tCT$, as well as the vortices $B_\pm$ which have $N_+ = 1$.

\end{document}